\newif\ifsinglecol
\singlecolfalse 

\ifsinglecol
	\documentclass[onecolumn, draftclsnofoot, journal]{IEEEtran}
\else
	\documentclass[10pt, twocolumn, final, twoside, journal]{IEEEtran}
\fi

\ifCLASSINFOpdf
   \usepackage[pdftex]{graphicx}
\else
\fi
\usepackage{array}
\usepackage{amsmath}
\usepackage{subcaption}
\usepackage{multirow}

\newcommand{\VW}{V}

\begin{document}
%
\title{Estimating snow cover\\ from publicly available images}
%
%
%

\author{Roman~Fedorov,~\IEEEmembership{Student Member,~IEEE,}
        Alessandro~Camerada,
		Piero~Fraternali,~\IEEEmembership{Member,~IEEE,}
		and~Marco~Tagliasacchi,~\IEEEmembership{Member,~IEEE}
\thanks{The authors are with the Dipartimento di Elettronica, Informazione e Bioingegneria, Politecnico di Milano, P.zza Leonardo da Vinci, 32 - 20133 Milano (Italy) e-mail: firstname.lastname@polimi.it}
\thanks{This work is supported by POR-FESR 2007-2013 PROACTIVE Project,
http://www.proactiveproject.eu}
\thanks{This work is supported by EU FP7 CUbRIK Integrating Project,
http://www.cubrikproject.eu}
}

\maketitle

\begin{abstract}
In this paper we study the problem of estimating snow cover in mountainous regions, that is, the spatial extent of the earth surface covered by snow. We argue that publicly available visual content, in the form of user generated photographs and image feeds from outdoor webcams, can both be leveraged as additional measurement sources, complementing existing ground, satellite and airborne sensor data. To this end, we describe two content acquisition and processing pipelines that are tailored to such sources, addressing the specific challenges posed by each of them, e.g., identifying the mountain peaks, filtering out images taken in bad weather conditions, handling varying illumination conditions. The final outcome is summarized in a snow cover index, which indicates for a specific mountain and day of the year, the fraction of visible area covered by snow, possibly at different elevations. We created a manually labelled dataset to assess the accuracy of the image snow covered area estimation, achieving 90.0\% precision at 91.1\% recall. In addition, we show that seasonal trends related to air temperature are captured by the snow cover index. 
\end{abstract}


%
\IEEEpeerreviewmaketitle

\section{Introduction}
Environmental monitoring requires collecting measurements of a very diversified range of physical quantities, which are then fed to models aimed at understanding past observations (e.g., climate change), detecting critical events in real-time (e.g., bush fires), and making predictions for the future (e.g., availability of water resources). Traditionally, such measurements are obtained by means of ad-hoc instrumentation that is designed, installed and managed by researchers and professionals interested in their analysis. 

The unprecedented availability of user generated data on the Web, and in particular in social media, poses unique opportunities for extracting valuable measurements from such data. These can be used to enrich traditional measurements by increasing coverage along both the spatial and temporal dimension. For example,~\cite{bradley2011outdoorwebcams} surveys the suitability of webcams as a new form of remote sensing for studying plant phenology (e.g., leaf growth and colour change), meteorology (e.g., cloud tracking, fog and visibility estimation, rainfall characterization, etc.), ethology (e.g., detecting bears), etc.

In this paper we are specifically interested in monitoring snow cover in mountainous regions, that is, the spatial extent of earth surface covered by snow. Snow processes are traditionally observed by means of ground measurements stations, which can either be manned or fully automated. In this case, measurements are accurate and capture different aspects, including the snow depth and density (possibly at different altitudes). However, the number of ground measurement stations is limited (for example, only $46$ stations are currently deployed over an area of $10500$ square kilometers covering the Italian Alps in the region of Lombardy), thus enabling only a sparse sampling of the snow cover over large areas. Moreover, the high variability of snow processes, which depend on temperature, elevation, exposure, slope, winds, etc., is such that it is difficult to extrapolate snow depth and density at different locations. An alternative source of measurements is represented by remote sensing, which relies on satellite~\cite{hall2002modis} or airborne~\cite{nolin1993estimating} imagery, synthetic aperture radar interferometry~\cite{shi1994sar}, laser scanner altimetry~\cite{prokop2008laseraltimetry}. These methods can provide a very high spatial coverage at moderate spatial resolution, but observations might not be available on a daily basis due to cloud cover and limited temporal frequency of satellite imagery. 

We explore the feasibility of leveraging user generated data to monitor snow cover. The objective is not to replace the use of ground-, satellite- or airborne-based measurements. Conversely, we argue that user generated data might represent an additional source that can complement and enrich existing ones, due to its unique characteristics in terms of spatio-temporal coverage and density, cost, etc. Specifically, we focus on \emph{visual} content and its accompanying metadata, which can be obtained from two different sources: user generated photographs posted on social media and image feeds from outdoor webcams. These sources have complementary characteristics. On the one hand, photographs are taken from different locations, possibly capturing different views of the same mountain peak, but their density varies significantly depending on the location (with higher spatial density near popular touristic destinations) and time of the year (with higher temporal density during holidays). On the other hand, webcams capture the very same view at a high temporal resolution. Although webcams are far more numerous than ground-based stations, they monitor a specific location and do not extensively cover large areas. 

Due to the distinct characteristics of photographs and webcams, we address them separately, designing two visual content processing pipelines tailored to the specific challenges posed by each source. To this end, we identify and retain for further analysis only those images depicting a mountainous landscape taken in good weather conditions (i.e., without occlusions due to clouds), for which it is possible to determine the location and the pose of the shot, so as to automatically identify the mountain peaks. For both photographs and webcam images we compute a \emph{snow cover index}, which indicates, for a specific mountain and day of the year, the fraction of the visible area covered by snow. Since we are able to align the image to the terrain, the snow cover index can also be disaggregated at different elevation ranges, so as to estimate the elevation of the snow line, i.e., the elevation at which the transition between snow and no-snow conditions is observed. 

The main contributions of the paper are as follows:
\begin{itemize}
	\item We design a content acquisition pipeline for user generated photographs posted on Flickr (see Section~\ref{sec:photos_acquisition}). Specifically, we crawled publicly available photographs and trained a classifier to distinguish images depicting mountainous landscapes from other images. We obtained a level of accuracy above 95\% on a balanced dataset containing more than $2k$ images by using state-of-the-art methods for image classification. 

	\item We propose a method to identify the mountain peaks appearing in a photograph (see Section~\ref{sec:photos_processing}). We exploited the availability of coarse-grained digital elevation models (DEMs) to perform the alignment between a photograph and a synthetic panorama rendered from the DEM. Knowing the position of the mountain peaks on the rendered panorama and the right alignment between the panorama and the photograph we are able to estimate the mountain peak positions present in the photograph. We show that it is possible to correctly identify the peaks in 75\% of the cases. This result increases to 81\% when considering only photographs with mountain slopes far from the observer. 
				
	 \item We describe a content acquisition and processing pipeline for image feeds from outdoor webcams (see Section~\ref{sec:webcams}). Specifically, we propose a method to automatically filter out images taken in bad weather conditions and aggregate images taken during the same day into a single representative image, removing transient cloud cover and compensating for time-varying shading.
	
	\item We compare different methods to automatically label each pixel of an image as representing terrain covered by snow or not (see Section~\ref{sec:snow_cover}). We tested them on a dataset for which ground-truth labels were available at the pixel level. We show that some of the methods work well for both user generated photographs and webcam images, achieving precision and recall values above 90\%.
	
	\item We define the snow cover index, show how seasonal trends related to air temperature and snow line are captured by its variation (see Section~\ref{sec:snow_cover_index}), and discuss how the snow cover index could be exploited to improve the estimation of the Snow Water Equivalent (SWE), a physical quantity of significant importance in environmental modeling (see Section~\ref{sec:conclusions}). 	
\end{itemize}

\section{Related Work}
In the past literature, several works addressed the study of snow processes by means of the analysis of visual content. The problems studied included estimation of glacier velocity, snow cover, depth and accumulation, and monitoring of snowfall processes. However, to the best of our knowledge, all works rely on one or more cameras designed and positioned ad hoc by researchers~\cite{laffly2012high}\cite{debeer2009modelling}\cite{garvelmann2013observation}\cite{hinkler2002automatic}\cite{parajka2012potential}. In most cases, the camera is set up so that specific objects (flags, sticks, etc.) are included in the field of view to ease calibration of color and/or geometry~\cite{laffly2012high}\cite{garvelmann2013observation}\cite{floyd2008measuring}\cite{parajka2012potential}. In addition, human assistance is needed to perform photo-to-terrain alignment~\cite{farinotti2010snow}\cite{debeer2009modelling}\cite{hinkler2002automatic} and to label pixels representing terrain covered by snow~\cite{laffly2012high}. Due to these constraint, the aforementioned works are not suitable to the scenario addressed in this paper, in which a very large number of images are collected in uncontrolled conditions. 

\subsection{Snow monitoring}
Some of the works adopt a single camera, purposely positioned and calibrated by the authors. Farinotti~et~al.~\cite{farinotti2010snow} combined melt-out patterns extracted from oblique photography with a temperature index melt model and a simple accumulation model to infer the snow accumulation distribution of a small Swiss Alpine catchment. However, the whole image processing pipeline was completely manual. It included choosing the photographs with the best meteorological and visibility conditions, photo-to-terrain alignment and snow covered area identification.
DeBeer~et~al.~\cite{debeer2009modelling} examined the spatial variability in areal depletion of the snow cover over a small alpine cirque of the Canadian Rocky Mountains, by observing oblique terrestrial photography. The images, obtained from a single ad-hoc installed digital high-precision camera, were projected on an extremely precise $1m$ resolution DEM. The orientation parameters were found manually for each image. The pixel level snow classification was obtained by means of a fixed threshold. This was possible because images were taken in short range, so that snow and terrain could be easily distinguished based on brightness alone. 
A similar problem was addressed by Hinkler et al.~\cite{hinkler2002automatic}, in which the authors derived snow depletion curves by projecting photographs obtained from a single ad-hoc camera onto the DEM. In this case, though, pixel-level labeling of snow was performed automatically exploiting RGB color components.

Other works adopt multiple cameras, which are positioned by the authors to monitor a specific area of interest. Laffly et al.~\cite{laffly2012high} combined oblique view ground-based pictures together with satellite images to produce a high temporal resolution monitoring of snow cover. The experiments were performed in the basin of a small polar glacier in Norway ($10 km^2$), with 10 digital cameras producing 3 images per day. The described method required a manual installation of $2 m \times 2 m$ orange flags on the snow at regularly spaced intervals to provide artificial reference points for photo-to-satellite matching. The identification of snow covered areas on the images was performed manually.
Garvelmann et al.~\cite{garvelmann2013observation} exploited a network of 45 spatially distributed cameras to obtain measurements of snow depth, albedo and interception in a German mountain range. Even if the results are highly correlated with ground-truth data, the proposed approach required the installation of wooden measurement sticks with alternating bars and plastic boards for compensating the different illumination conditions of each camera.
Parajka et al.~\cite{parajka2012potential} investigated the benefit of using terrestrial photography for both short-range and far-range views. In case of short-range, analysis measurement sticks were installed in front of the cameras, whereas in the far-range the authors did not identify snow, but simply compared the photographs with simulations of snow distribution.
Floyd~et~al.~\cite{floyd2008measuring} monitored the snow accumulation during the rain-no-snow events by means of the acquisition of photographs from cameras designed and positioned ad-hoc. This approach required the installation of measurement sticks within the camera field of view. The analysis was performed on a short-range view, so that a fixed pixel intensity threshold was enough to perform pixel-level snow classification.

With respect to the aforementioned works, we propose processing pipelines that are able to scale to a large number of images collected from either user generated photographs or outdoor webcams. This requires the design of fully automated components to perform photo-to-terrain alignment and pixel-level snow classification. 
\subsection{Mountain peak identification}
Photo-to-terrain alignment was addressed by Baboud et al.~\cite{Baboud2011Alignment}. However, their method was not quantitatively evaluated on a large dataset, and qualitative results were provided for $28$ photographs only. The examples reported in the paper reveal a very accurate alignment with the terrain, which was possible thanks to the use of a high-resolution DEM, not available for all mountain regions.
Baatz et al.~\cite{Baatz:2012:LSV:2403006.2403045} approached a related problem, that is, the estimation of the geographical position of mountain photographs in the absence of geo-tags by means of content based analysis. However, they did not address how to determine the labels of the mountain peaks. In addition, in some of the examples, the sky-to-terrain segmentation was performed manually, before the photograph was processed by the algorithm.
Liu and Su~\cite{Liu:peakRecognition} presented a content-based image search method based on the shape of the skyline. The idea is to match two photographs that contain the same peaks, similarly to landmark search in urban environments. However, photo-to-terrain alignment and labeling of mountain peaks is not supported.
The proposed algorithm for photo-to-terrain alignment described in Section~\ref{sec:photos_processing} was preliminary discussed by the authors in their previous work~\cite{fedorov2013exploiting} and recently extended in~\cite{fedorov2014mountain}. Unlike~\cite{Baboud2011Alignment}, we provide a quantitative evaluation on a significantly larger dataset and introduce different adjustments in the preprocessing and alignment algorithm, needed to cope with photos taken in diverse weather conditions and in the presence of other objects (trees, mountain slopes in the foreground, etc.). In addition, we adopt a coarse resolution DEM, which is publicly available for the whole earth surface. Conversely,~\cite{Baatz:2012:LSV:2403006.2403045} is based on an extremely precise DEM available only for Switzerland ($swissALTI^{3D}$: $2m$ spatial resolution), and it is not obvious how similar results can be achieved in a different area.

\subsection{Pixel-wise snow classification}
The problem of automatically detecting the presence of snow at the pixel-level was addressed in just a few works. As mentioned above, both~\cite{debeer2009modelling} and~\cite{floyd2008measuring} perform a simple thresholding of brightness values. However, this is applicable only to short-range views. Full color information was exploited in~\cite{hinkler2002automatic}, which proposed a snow index based on a normalized difference between RGB components. Similarly, ~\cite{salvatori2011snow} presented a simple algorithm for pixel-level snow classification based on thresholding the blue color component, in which the threshold is determined automatically based on the statistical analysis of the image histogram. The method produced excellent results (precision above $0.99$), but was tested in somewhat controlled conditions, with short-range views without shadows and cloud occlusions. More recently, R{\"u}fenacht et al.~\cite{rufenacht2014temporally} proposed a method based on Gaussiam-Mixture-Model (GMM) clustering of RGB pixel values, designed to work for long-range images of mountain slopes. All these methods (\cite{hinkler2002automatic},~\cite{salvatori2011snow},~\cite{rufenacht2014temporally}) are included in the experimental evaluation in Section~\ref{sec:snow_cover}. 
\begin{figure*}[t]
\centerline{\includegraphics[width=\textwidth]{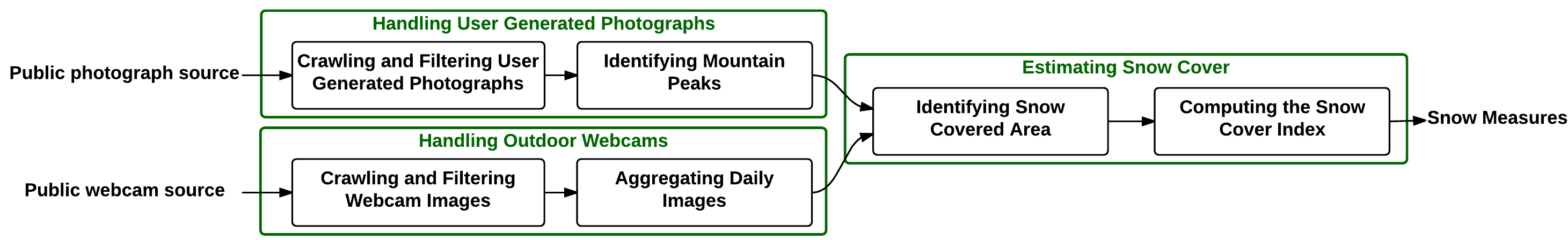}}
\caption{A schematic overview of content processing pipelines to estimate snow cover.}
\label{fig:pipeline}
\end{figure*}

\section{Handling user generated photographs}\label{sec:photos}
A very large number of user generated photographs are publicly available online. However, in order to be of any use for the purpose of estimating snow cover, two issues must be properly addressed by the proposed processing pipeline, as illustrated in Figure~\ref{fig:pipeline}. First, photographs need to be crawled and filtered, so that only relevant photographs (i.e., those depicting mountain landscapes) are retained. This is addressed in Section~\ref{sec:photos_acquisition}, which describes an automatic image classification algorithm for this purpose. Second, once a photograph is deemed relevant, it must be aligned to the terrain, so that we can unambiguously determine the mountain peaks appearing in the image. This is covered in Section~\ref{sec:photos_processing}, which describes an automatic photo-to-terrain alignment algorithm. 

\subsection{Crawling and filtering user generated photographs}\label{sec:photos_acquisition}
Flickr was selected as the data source for user generated photographs, because it contains a large number of publicly available images, many of which have an associated geo-tag (GPS latitude and longitude position saved in the EXIF container of the photograph). We crawled a $300 \times 160$ kilometer region across the Italian and Swiss Alps (in the area of Pennine Alps, Lepontine Alps, Rhaetian Alps and Lombard Prealps). By using the Flickr API it is possible to query the service using temporal and spatial filters. However, each query is limited to return a maximum of 4000 records. The algorithm is designed to start from the whole region of interest and recursively split it into subregions and then perform separate queries. This is performed until the sub-regions have an image count (information provided by the API) lower than the maximum allowed, so as to retrieve all the publicly available images in the desired area.
This resulted in 600k photographs with a valid geo-tag in the temporal window between January 2010 and July 2014. Then, in order to determine the elevation at which the shot was taken, we employed the publicly available Shuttle Radar Topography Mission (SRTM3) DEM, which provides the elevation for points spaced 3 arcseconds apart, approximately corresponding to a spatial resolution of 90 meters. 

\begin{figure}[t]
	\centering
        \centerline{\includegraphics[width=0.5\textwidth]{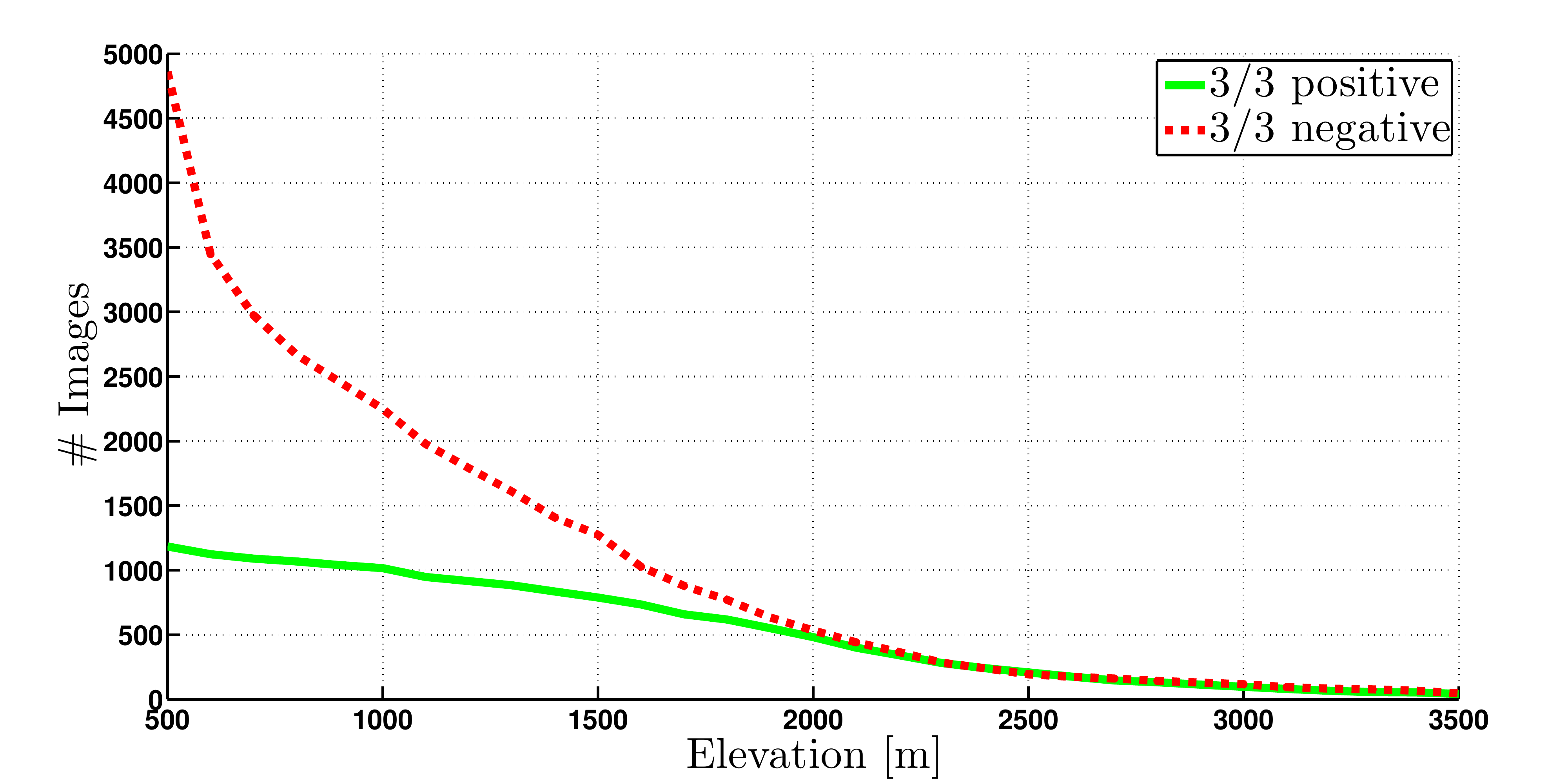}}
	\caption{Histogram of the number of positive/negative samples at different elevations.}
	\label{fig:crowdsourcing_histogram}
\end{figure}

In order to train a classifier to identify photographs of mountain landscapes, we collected ground-truth labels from a subset of the crawled images. 6940 randomly selected photographs from those taken above 500 meters were processed in a crowdsourcing experiment\footnote{We used the Microtask platform - http://microtask.com}, designed to collect three labels for each photograph. Specifically, each worker was asked to label each image by answering to the following question: ``Does this image contain a meaningful skyline of a mountain landscape?''. In order to clarify the expected outcome of the task, we provided a tutorial with some selected images representing both positive (mountain) and negative (no-mountain) samples. The labeling task required approximately 1 second per image. The aggregated label was then obtained by means of majority voting. The results of the crowdsourcing experiment are reported in Table~\ref{tab:crowdsourcingHist}. Approximately 23\% (17.0\% + 6.1\%) of the images were classified as positive. Note that in almost 13\% of the cases there was not full agreement among workers, due to the subjective nature of the task. Figure~\ref{fig:crowdsourcing_histogram} illustrates the number of positive/negative images for each elevation range. 50\% of the images taken above 2000 meters represent mountain landscapes and the number of negatives rapidly grows below 600 meters. Hence, we retained only the images above such elevation for further analysis, ending up with a total of 237k images.

\begin{table}[t]
	\begin{center}
	\caption{Results of the crowdsourcing experiment}
	\label{tab:crowdsourcingHist}	
	\begin{tabular}{|l|r|r|}
		\hline
		outcome & percentage & count\\
		\hline
		3/3 positive & 17\% & 1184\\
		2/3 positive & 6.1\% & 422\\
		2/3 negative & 7\% & 483\\
		3/3 negative & 69.9\% & 4851\\
		\hline
	\end{tabular}
	\end{center}
\end{table}

\textbf{Feature extraction and encoding}: from each of these images we computed a fixed-dimensional feature vector which summarizes its visual content. Following the experiments proposed by Xiao et al.~\cite{xiao2014sun} (and partially using their code available online) we evaluated the classifier performance with several local and global feature descriptors.

\emph{Dense SIFT}: we extracted SIFT descriptors from color images by testing different color models at different scales $\{\frac{2}{3},1,\frac{4}{3},\frac{5}{3}\}$, sampled from a uniformly spaced grid with step size equal to $6\times6$ pixels, obtaining around $10^5$ descriptors for each image (the exact number depends on the image resolution). The descriptors for RGB, HSV and opponent color models were obtained as the concatenation of the SIFT descriptors of each color channel.
The Bag-of-Visual-Word (BoVW) model was adopted, encoding the feature vector as a histogram of visual words with a dictionary determined during an off-line training phase. Specifically, $10^2 \cdot \VW$ SIFT descriptors were randomly sampled from 100 randomly selected images, where $\VW$ denotes the number of visual words in the dictionary. The dictionary was learned using $k$-means, with $k = \VW$. 

\emph{HOG2x2}: as with Dense SIFT, Histogram of Oriented Edges (HOG) descriptors were densely extracted, computing a histogram of oriented gradients in each $8 \times 8$ pixels cell and normalizing the result using a block-wise pattern (with $2 \times 2$ square HOG blocks for normalization). We adopted UoCTTI HOG variant~\cite{felzenszwalb2010object}. Similarly to Dense SIFT, the BoVW model was adopted.

\emph{SSIM}: self-similarity descriptors~\cite{shechtman2007matching} were computed on a regular grid at $5 \times 5$ pixels step. Each descriptor was obtained as a correlation map of a patch of $5 \times 5$ in a window with radius equal to $40$ pixels, quantified in $3$ radial bins and $10$ angular bins. Similarly to Dense SIFT, the BoVW model was adopted.

\emph{GIST}: the GIST descriptor~\cite{oliva2001modeling} was computed as a wavelet image decomposition (each image location is represented by the output of filters tuned to different orientations and scales). We adopted the parameter setting proposed in~\cite{xiao2014sun}. The result was a global image descriptor of $512$ dimensions.


We also explored the effects of replacing the BoVW model with the Fisher Vector encoding as described in~\cite{sanchez2013image}. We studied the Fisher Vector encoding applied with different number of Gaussians, and with/without PCA.

In order to capture spatial clues, we adopted the spatial histogram approach proposed by~\cite{grauman2005pyramid} and~\cite{lazebnik2006beyond}. 
In addition to computing a Dense SIFT, HOG2x2 or SSIM $\VW$-dimensional histogram for the whole image, we also split the image in three equally sized horizontal tiles, and computed a $\VW$-dimensional histogram for each tile. Each of the four histograms (total and three tiles) was $L_1$-normalized and then stacked to form a $4\VW$-dimensional vector, which was $L_2$-normalized.
Analogously to spatial histogram approach for local features, GIST descriptors were extracted from the whole image and three images representing equally sized horizontal tiles, then concatenated.
A similar technique was applied in case of the Fisher Vector, concatenating four encodings (all image features and one for each horizontal tile).

\textbf{Experiments}: the feature vectors were fed to a SVM classifier using a $\chi^2$ kernel. 
In order to create a balanced dataset, we retained all the 1184 positive samples and randomly selected the same number of negative samples. Then, we used 1658 samples ($\sim$70\%) for training and validation and 710 samples ($\sim$30\%) for testing. In order to learn the optimal values of the parameters of the SVM classifier, we adopted $k$-fold cross validation, with $k = 5$. Thus, the set of labelled samples for training and validation was split in $k$ disjoint sets. At each iteration, one set was used for validation, while the others were used for training. We performed a grid search to seek the optimal hyper-parameters $C$ and $\gamma$ of the kernel, each parameter in the set $\{0.01, 0.033, 0.066,0.1.0.33,0.66,1,3.3,6.6,10,33,66,100\}$. 

\textbf{Results}:
Table~\ref{tab:image_classification_results_allfeatures} summarizes the results obtained within the test set, by all listed feature extractors, with the best configuration in terms of $C$ and $\gamma$ of the SVM kernel. Performance was measured using accuracy, defined as the fraction of samples for which the classifier provides the correct label. For completeness, Table~\ref{tab:image_classification_results_allfeatures} also shows the values of precision and recall.
HOG2x2 obtains similar performance to Dense SIFT; both Dense SIFT and HOG2x2 slightly outperform SSIM. All three local feature descriptors (Dense SIFT, HOG2x2, SSIM) perform better than GIST.

Table~\ref{tab:image_classification_results} shows the detailed results obtained by Dense SIFT (as the best performing feature) within the test set, for different sizes of the dictionary $\VW \in \{2500, 5000\}$ and for each color model. In all cases, we obtained very good results, with the highest value of accuracy (above 95\%) achieved by using the RGB color model, regardless of the number of visual words adopted. We also computed the learning curves indicating the accuracy for both the training and the test set, to exclude overfitting and verify that no additional gains could be expected by further increasing the size of the training set. In addition, although not shown in Table~\ref{tab:image_classification_results}, we investigated the use of different vocabulary sizes (namely $\VW = 1000$ and $\VW = 10000$), which did not improve the accuracy. Furthermore, we investigated the effect of the replacement of the BoVW model with the Fisher Vector encoding. We used different number of Gaussians for the Fisher Vector (namely $16$ and $128$), with and without applying PCA. None of the configurations of the Fisher Vector improved the accuracy.

Finally, the images that are classified as positive are passed to the next step of the pipeline, to identify the mountain peaks. This phase is described in the next section.

\begin{table}[t]
	\begin{center}
		\caption{Results obtained by different feature extractors for the image classification problem (mountain vs. no-mountain).}
		\label{tab:image_classification_results_allfeatures}
	\begin{tabular}{|c|c|c|c|c|c|}
		\hline
		feature & $C$ & $\gamma$ & accuracy & precision & recall\\
		\hline
		Dense SIFT & 3.3 & 0.66 & \textbf{95.1} & \textbf{94.0} & \textbf{96.3}\\
		HOG2x2 & 3.3 & 0.033 & 94.7 & 93.9 & 95.5\\		
		SSIM & 0.66 & 0.33 & 93.0 & 92.5 & 93.5\\		
		GIST & 0.33 & 1 & 87.61 & 82.64 & 95.21\\
		\hline
	\end{tabular}
	\end{center}
\end{table}

\begin{table}[t]
	\begin{center}
		\caption{Results obtained by Dense SIFT for the image classification problem (mountain vs. no-mountain).}
		\label{tab:image_classification_results}
	\begin{tabular}{|c|c|c|c|c|c|c|}
		\hline
		model & $\VW$ & $C$ & $\gamma$ & accuracy & precision & recall\\
		\hline
		gray & 2500 & 3.3 & 0.1 & 93.6 & 91.9& 95.8\\
		gray & 5000 & 1 & 0.33 & 94.4 & 93.6 & 95.2\\
		RGB & 2500 & 0.33 & 0.01 & \textbf{95.1} & \textbf{94.7} & 95.5\\
		RGB & 5000 & 3.3 & 0.66 & \textbf{95.1} & 94.0 & 96.3\\
		HSV & 2500 & 33 & 0.66 & 94.2 & 92.0 & \textbf{96.9}\\
		HSV & 5000 & 6.6 & 1 & 94.1 & 92.9 & 95.5\\
		opponent & 2500 & 0.66 & 0.66 & 94.0 & 92.0 & 96.6\\
		opponent & 5000 & 1 & 0.33 & 93.2 & 90.7 & 96.3\\		
		\hline
	\end{tabular}
	\end{center}
\end{table}

\subsection{Identifying mountain peaks}\label{sec:photos_processing}
\begin{figure*}
\centerline{\includegraphics[width=\textwidth]{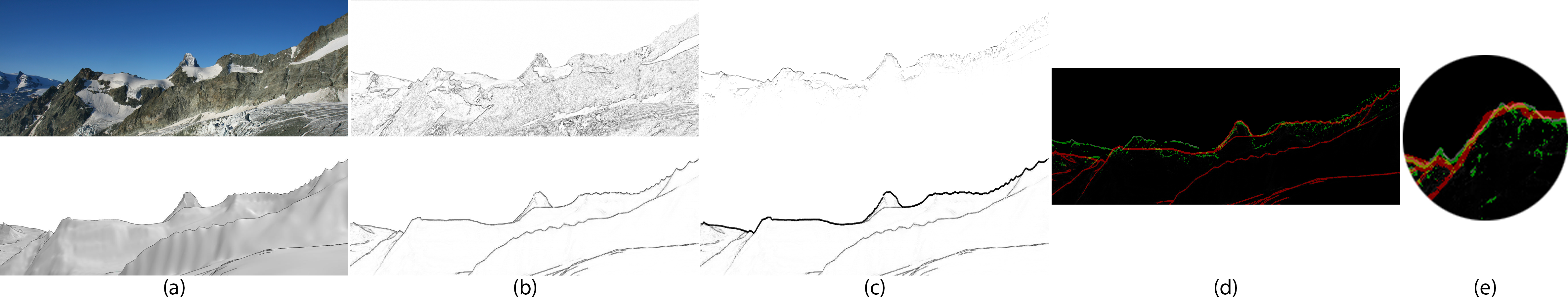}}
\caption{An example of the photo-to-terrain alignment: (a) input photograph (top) and corresponding panorama (bottom), (b) edge extraction, (c) skyline detection, filtering and dilation (d) global alignment with refinement (e) local alignment.}
\label{fig:peakIdentificationSchema}
\end{figure*}

After a user-generated photograph is recognised to contain a mountain landscape, it is processed to align it with the terrain. To this end, given a photograph and the metadata extracted from the EXIF container (geo-tag, focal length, camera model and manufacturer), it is possible to perform a matching with a $360^\circ$ panoramic view of the terrain synthesized from a DEM. The alignment method proceeds in four steps, described below and illustrated in Figure~\ref{fig:peakIdentificationSchema}.

\emph{Preprocessing}: The horizontal Field Of View (FOV) of the photograph is calculated from the focal length and the size of the camera sensor:
$$
		FOV = 2\arctan \frac{w}{2f},
$$
	where $f$ is the focal length of the camera and $w$ is the width of the sensor. Then, the photograph is rescaled considering that the width of the panorama corresponds to a FOV equal to $360^\circ$. This is necessary to ensure that the photograph and the panorama have the same scale in degrees per pixel and thus matching can be performed without the need of scale invariant methods. The EXIF metadata provide the focal length $f$, but not always the sensor specifications of the camera. Therefore, the width of the sensor is obtained by querying a publicly available database, which contains technical specifications for a very large number of consumer camera models\footnote{http://www.digicamdb.com}.

	Due to the different nature of the photograph and of the panorama (Figure~\ref{fig:peakIdentificationSchema}a), it is not possible to exploit conventional descriptors, e.g., color, texture or local features. However, edges can be employed to match the images. Hence, we apply an edge extraction algorithm to both the photograph and the panorama to produce an edge map, which assigns to each pixel the strength of the edge at that point and its direction (Figure~\ref{fig:peakIdentificationSchema}b).
We use the Compass edge detector~\cite{conf/cvpr/RuzonT99}, which performs well also within regions with similar color and texture. This is a common situation in photographs of mountain landscapes, e.g., whenever a snow-capped peak is in front of a bright sky background. Compass is particularly robust in detecting edges corresponding to the skyline, which is the most important visual feature exploited by the alignment algorithm. 

	Matching edges of an image with those of a synthetic panorama requires addressing the fact that there is not a one-to-one mapping between edge pixels extracted from the two images. In particular, the edges extracted from the photograph include several noisy edge pixels that do not correspond to any orografic feature of the mountain slopes, but to other objects in the foreground (e.g., rocks, trees, lakes, houses, etc.) and in the background (e.g., clouds, snow patches, etc.).
This calls for a method to identify the few, yet discriminative, edge pixels that can be reliably used to align the photograph to the panorama. This can be accomplished in two steps: first, a skyline detection algorithm is employed~\cite{Lie2005221}, and all the edge pixels above the skyline are removed, being considered obstacles or clouds. Second, a simple weighting mechanism is applied, which assigns decreasing weights to the edge pixels as the distance from the skyline increases (Figure~\ref{fig:peakIdentificationSchema}c - top).


	As for the panorama, the edges corresponding to the skyline can be simply identified as the upper envelope of the edge map, by keeping, for each column of pixels, the topmost edge point. Since the edge filtering of the photograph emphasize the edges of the skyline, a morphological dilation is applied to emphasize the edges corresponding the skyline of the panorama (Figure~\ref{fig:peakIdentificationSchema}c - bottom).


\emph{Global alignment}: The matching between the photograph and the corresponding panorama is performed using a Vector Cross Correlation (VCC) technique, also used in~\cite{Baboud2011Alignment}, which takes into account both the strength and the direction of the edge points. The output of the VCC is a correlation map that, for each possible horizontal and vertical displacement between the photograph and the panorama, indicates the strength of the matching. Then, the top-$K$ local maxima of the correlation map are identified as candidate matches.

	Global alignment can match mountain edges also below the skyline and is robust with respect to skyline detection errors. However, the global maximum of the correlation is not necessarily the correct match. This might occur, for example, when some edges of the photograph happen to match the shape of different portions of the panorama. As such, the top-$K$ matches are further analyzed by the refinement step below.

\emph{Refining global alignment}: For each of the top-$K$ candidate matches, we measure the Hausdorff distance between the skyline edge points of the photo and of the panorama, when the two are overlapped at the candidate matching position. A scoring function is computed, which combines the Hausdorff distance and the rank position computed by the initial global alignment. The candidate with the highest score is then chosen as the best match between the photo and the panorama (Figure~\ref{fig:peakIdentificationSchema}d).
	
\emph{Local alignment}: Our method generates a panorama from a coarse DEM, using a possibly noisy geo-tag. Therefore, in most cases the panorama does not match the photo perfectly, thus increasing the difficulty of finding a correct global alignment.
Therefore, to improve the precision of the position of each mountain peak, a separate VCC procedure is applied, similar to the one used in the global alignment step.
Specifically, for each peak we consider a local neighborhood centered in the photograph location identified as the peak position by the global alignment. In this way each peak position is refined by identifying the best match in its local neighborhood.
Overall, this is equivalent to applying a non-rigid warping of the photograph with respect to the panorama.



\textbf{Results}: Our method was tested on a set of photographs selected from those crawled with the method described in Section~\ref{sec:photos_acquisition}. We manually inspected a subset of 200 photographs and the panoramas generated based on the accompanying EXIF metadata to make sure that a plausible matching existed. Indeed, in some cases, we found that the geo-tag was available but incorrect, such that the generated panorama could not be matched to the photograph by any means. Ultimately, we retained 162 photographs in our test set.
Then, the ground truth data was generated by an alignment tool developed ad-hoc,
which allows the user to find the correct position of the photograph in the panorama and then to locally warp the image by overlapping each mountain peak present in the photo to the corresponding one in the panorama. 

For each peak $i = 1, \ldots n$, let $(x^p_i, y^p_i)$ and $(x^r_i, y^r_i)$, denote the pixel coordinates in the coordinate system of the photo and of the panorama, respectively. When the photo is aligned with a displacement $(\Delta x,\Delta y)$, we define the angular error in the position of the $i$-th peak as
$$\epsilon_i(\Delta x,\Delta y) = \sqrt{d_x(x^r_i,\Delta x+x^p_i)^2+d_y(y^r_i,\Delta y+y^p_i)^2},$$

where $$d_x(x_1,x_2) = (360/w_r)\min(w_r - |x_1-x_2|,|x_1-x_2|),$$ is the angular distance (in degrees) between two points along the azimuth, given the circular symmetry of the panorama,  and $w_r$ is the number of pixels corresponding to $360^\circ$. Similarly $$d_y(y_1,y_2) = (360/w_r)|y_1-y_2|,$$ where the same angular resolution in degrees/pixel is assumed due to small elevation angles. When creating the ground truth, the images are warped so as to minimize the average angular error $$\epsilon(\Delta x,\Delta y) = (1/n)\sum_{i = 1}^n \epsilon_i(\Delta x,\Delta y),$$ where $n$ is the number of peaks, and to find the best displacement $$(\Delta x^*, \Delta y^*)=\arg\min_{\Delta x,\Delta y}\epsilon(\Delta x,\Delta y).$$
Note that $\epsilon^* = \epsilon(\Delta x^*,\Delta y^*)$ cannot always be reduced to 0, due to the coarse granularity of the panorama.

Let $(\Delta x^G_k, \Delta y^G_k)$, $k = 1,\ldots, K$, denote the displacements of the top-$K$ candidate matches of global alignment. We define $p^G_{K}(\theta)$  as the fraction of the photos in the test set that have at least one candidate match displacement $(\Delta x^G_k, \Delta y^G_k)$ lying within angular distance $\theta$ from the ground truth $(\Delta x^*,\Delta y^*)$. The refinement step selects $(\Delta x^R_K, \Delta y^R_K)$ to be one of the displacements $(\Delta x^G_k, \Delta y^G_k)$ (not necessarily the best). Then, $p^R_K(\theta)$ is the fraction of photographs for which the difference between $(\Delta x^R_K, \Delta y^R_K)$ and $(\Delta x^*,\Delta y^*)$ is below $\theta$. Note that $p^R_{K}(\theta) \le p^G_{K}(\theta)$ by construction, and the equality holds if the refinement step is always able to identify the correct match within the top-$K$ candidates.

The local alignment step computes a different displacement $(\Delta x^L_i,\Delta y^L_i)$ for each of the $n$ peaks. Then, the average error is defined as $$\epsilon^L = (1/n)\sum_{i = 1}^n \epsilon_i(\Delta x^L_i,\Delta y^L_i).$$


\begin{figure}[t]
\centerline{\includegraphics[width=0.5\textwidth]{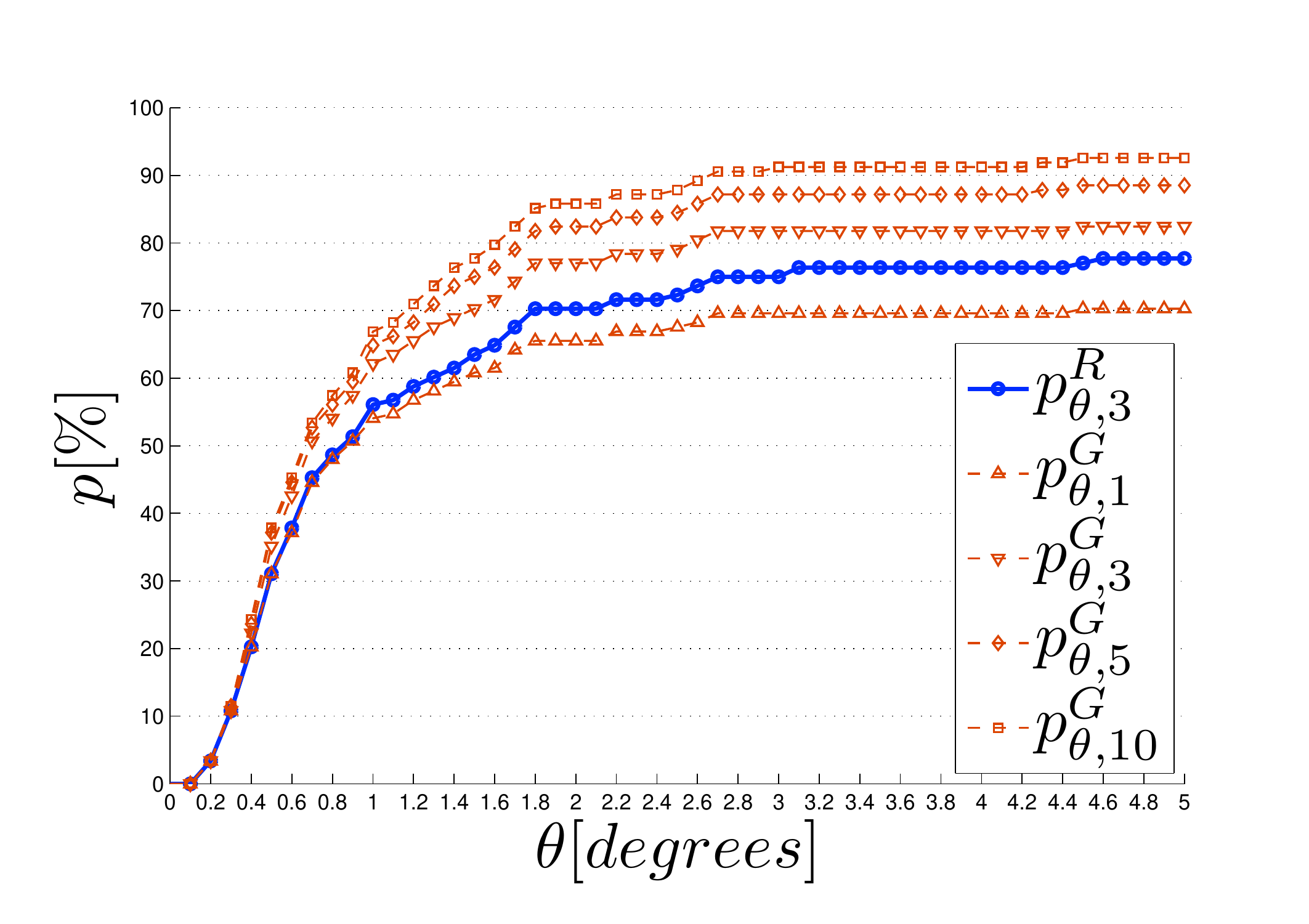}}
\caption{Performance analysis of the global alignment and of the refinement step.}
\label{fig:peakIdentificationResults}
\end{figure}

Figure \ref{fig:peakIdentificationResults} shows the performance of the global alignment procedure on the entire dataset. It can be observed that $p^G_{K}(\theta)$ saturates when $\theta$ exceeds $3^\circ$. Specifically, 69.6\% of the photographs are aligned with an average error below $3^\circ$, when considering the top-1  match. The fraction of  correctly aligned photos grows to 81.8\%, 87.2\% and 91.2\% when $K$ is 3, 5 and 10, respectively. Diminishing returns in the average error are observed when increasing $K$; thus, we selected $K = 3$ in the refinement step, which results in 78\%  of correctly aligned photos. The refinement performance curve lies approximately halfway between the top-1 and top-3 curves of global alignment. This shows the benefit of introducing the refinement step and its ability to pick the correct candidate from the top-3 candidates.

%

Taking a deeper look into the dataset, Table~\ref{tab:peakIdentificationResults} describes the performance of the proposed method depending on the different properties of the visual content, manually annotated in two ways; first, we marked whether the photograph contains clouds  (80 out of 162); second, we marked the presence of mountains close to the observer that occlude the skyline in the background (49 out of 162).
The presence of clouds is one of the main obstacles to be addressed. This is due to the fact that, when clouds partially occlude the skyline, the outcome of the skyline detection algorithm might fail. In addition, edge points due to clouds above the skyline might compromise the filtering procedure, which is based on the assumption that there are no edges above the skyline. In the case of global alignment, the fraction of correctly matched photographs grows to 72.4\% and 82.9\% in the absence of clouds, when considering the top-1 and top-3 candidates, respectively. Conversely, the presence of clouds leads to a reduction of correct matches, which represent, however, at least 66.7\% of the cases. The performance of the refinement step is also affected by the presence of clouds, being equal to 77.6\% (72.2\%) when clouds are absent (present). The impact of clouds is higher in the refinement step than in the top-3 candidates global alignment, because the former relies heavily on the correctness of the estimated skyline.
Another issue lies in the presence of mountain slopes nearby the observer. Indeed, in this case small errors in the geo-tag might lead to a panorama which does not correctly represent the viewpoint of the photograph.
In the case of global alignment, the fraction of correctly matched photographs grows to 74.8\% and 89.3\% in the absence of nearby mountains, when considering the top-1 and top-3 candidates, respectively. A similar behavior is observed for the refinement step (81.6\%).

\begin{table}[t]
\caption{Performance results of the photo-to-terrain alignment algorithm (by dataset categories and photograph content properties)}
\label{tab:peakIdentificationResults}
\setlength{\extrarowheight}{3pt}
\centering
\begin{tabular}{c|c|c|c|}
\cline{2-4} 
\multicolumn{1}{l|}{}                                 & $p^G_{3,1}$ & $p^G_{3,3}$ & $p^R_{3,
3}$ \\ \hline
\multicolumn{1}{|l|}{All images}              	  & 69.6\%        & 81.8\%         & 75.0\%           \\ \hline
\multicolumn{1}{|l|}{Absence of clouds}               & 72.4\%       & 82.9\%        & 77.6\%           \\
\multicolumn{1}{|l|}{Presence of clouds}              & 66.7\%       & 80.6\%       & 72.2\%           \\ \hline
\multicolumn{1}{|l|}{Absence of nearby mountains} 	  & 74.8\%       & 89.3\%       & 81.6\%           \\
\multicolumn{1}{|l|}{Presence of nearby mountains}    & 57.8\%       & 64.4\%       & 60.0\%           \\ \hline
\end{tabular}
\end{table}

Local alignment further improves the matching between the photograph and the panorama. This is measured by comparing the average angular error between the peak positions after the refinement step, $\epsilon(\Delta x_K^R,\Delta y_K^R)$, with the value obtained after local alignment, $\epsilon_L$. In our experiments, we found that the error decreased from $\epsilon(\Delta x_K^R,\Delta y_K^R) = 0.99^\circ$ to $\epsilon_L = 0.78^\circ$, i.e., a 21\% reduction.

Unfortunately it was not possible to compare our results with those obtained by other algorithms discussed earlier, due to the lack of a publicly available implementation of the method and unspecified quantitative evaluation metrics~\cite{Baboud2011Alignment}. Instead, \cite{Baatz:2012:LSV:2403006.2403045} and \cite{Liu:peakRecognition} address different problems (respectively, geo-tag estimation and relevant image retrieval) and cannot be compared directly with our work.

\section{Handling outdoor webcams}\label{sec:webcams}

\subsection{Crawling and filtering webcam images}\label{sec:webcams_acquisition}

Outdoor webcams represent an additional valuable source of visual content that can be exploited to monitor snow cover. The use of selected webcams that point to mountain landscapes poses different advantages and disadvantages with respect to user generated photographs. On the one hand, the images captured by a webcam need not to go through the relevance classification pipeline described in Section~\ref{sec:photos}, because they contain useful content and the alignment with the terrain model can be performed once, possibly with manual supervision. In addition, most webcams capture images every 1 to 15 minutes, thus ensuring a very high temporal density. On the other hand, the spatial density is lower than that of user generated photographs, because they tend to be deployed mostly near popular touristic destinations. Moreover, due to bad weather conditions that significantly affect short- and long-range visibility (e.g., clouds, heavy rains and snowfalls), only a fraction of the images can be exploited as a reliable source of information for estimating snow cover. In this respect, we manually screened $1000$ images crawled from 4 webcams (Valmalenco - Italy, Bormio - Italy, Metschalp - Switzerland, Hohsaas - Switzerland) in daytime hours (9:00-18:00)
and we observed that $67\%$ of them were not suitable for further analysis due to insufficient visibility. 

Therefore, we devise a simple algorithm that automatically filters out those images acquired during bad weather conditions. The key assumption is that, when visibility is sufficiently good, the skyline of the mountain profile is not occluded. For each webcam, we create a binary mask $L$ with the same size of the acquired image. Such binary mask indicates those pixels $p = (x,y)$ that are in the neighborhood of the skyline. Hence
$$
	L(p) = \left\{
	\begin{array}{ll}
		1 & \mbox{if }  \exists r \in \mathcal{L}: \| p - r\| \le \tau \\
		0 & \mbox{otherwise}\\
	\end{array}
	\right.,
$$
where $\mathcal{L}$ denotes the set of pixels that belong to the skyline, $\|\cdot\|$ computes the Euclidean norm. We empirically set $\tau = 0.04h$, where $h$ denotes the height of the image in pixels.
Then, for each image acquired by a webcam, we compute its edge map $E$ using the same method as in Section~\ref{sec:photos} and we binarize the result. We define a function $f(\cdot)$ that, given an image, returns the number of columns that contain at least one non-zero entry, and the skyline visibility score as
$$
	v = \frac{f(E \cdot L)}{f(L)},
$$	
where $\cdot$ denotes the pixel-wise product between two images of the same size. The value of $v$ is in the interval $[0,1]$ and can be intuitively interpreted as the fraction of the whole skyline that is visible in a given image. We retain for further processing only those images for which $v \geq \bar{v}$, where $\bar{v}$ is a threshold, which was set to $0.75$, based on the experiments illustrated below. The proposed method retains images in which clouds do not occlude the skyline, although they might still be present and interfere with estimating the snow cover. However, transient clouds are handled and removed by the method described Section~\ref{sec:webcams_processing}. 


\textbf{Results:}
In order to build a reliable test dataset, we manually labeled $1000$ images collected from $4$ webcams. Each image was manually tagged as ``good weather'', if the entire mountain area was visible and not occluded by clouds, or as ``bad weather'' otherwise.

\begin{figure}
\centerline{\includegraphics[width=0.5\textwidth]{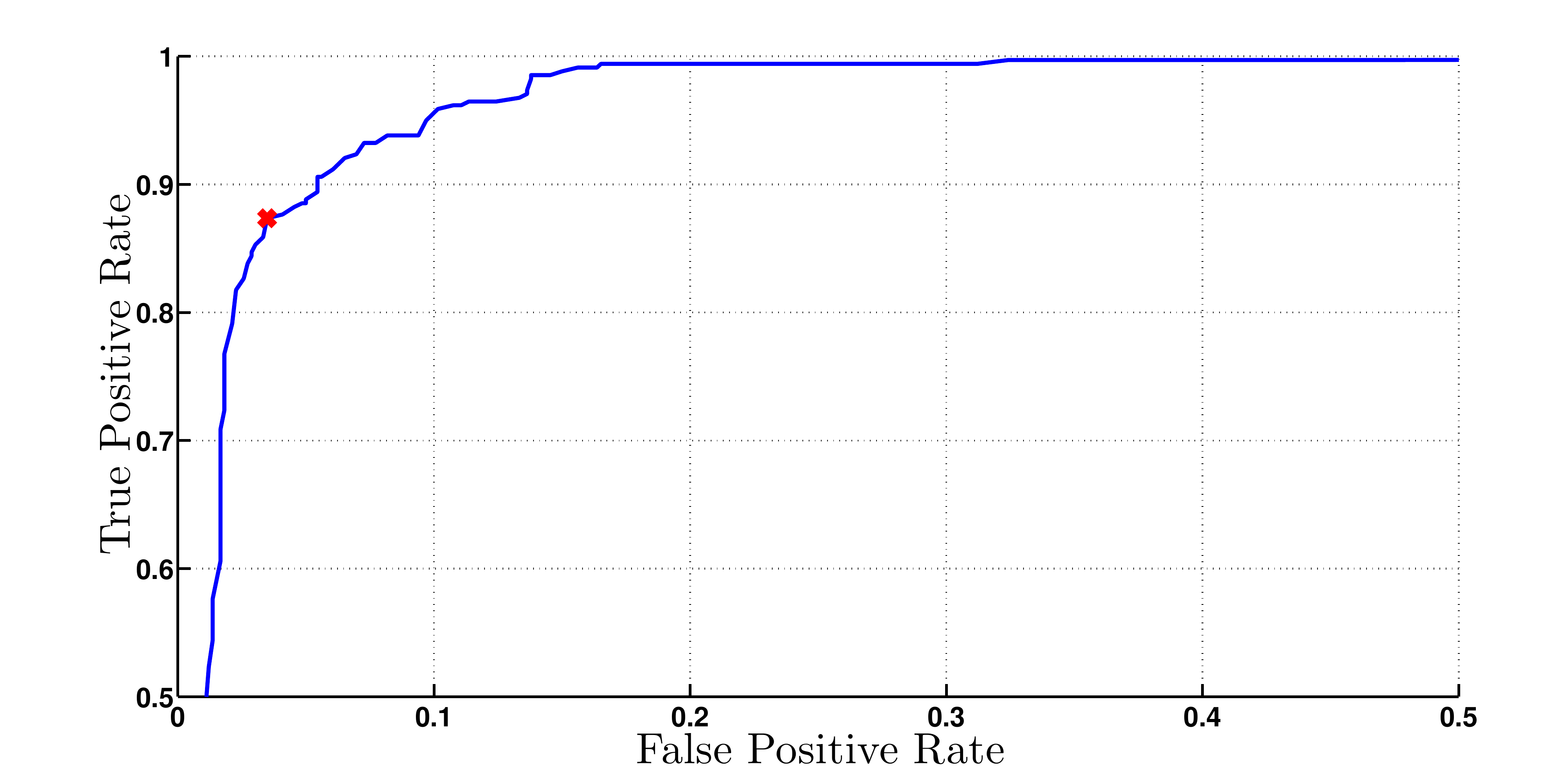}}
\caption{The ROC curve of the good weather webcam image classifier when varying the threshold $\bar{v}$.}
\label{fig:goodWeatherClassifierROC}
\end{figure}

The classifier was evaluated using a ROC curve, which shows the True Positive Rate (TPR) vs. the False Positive Rate (FPR), illustrated in Figure~\ref{fig:goodWeatherClassifierROC}. The temporal frequency of the webcam image acquisition is high, so a large number of images is available. Hence, the choice of the threshold parameter $\bar{v}$ was driven by the goal of having low FPR. Namely, $\bar{v}$ was set to $0.75$ (corresponding to the point marked in Figure~\ref{fig:goodWeatherClassifierROC}), obtaining a TPR equal to $87.4\%$ at FPR $3.5\%$.

\subsection{Aggregating daily images}\label{sec:webcams_processing}

\begin{figure*}
\centerline{\includegraphics[width=1.0\textwidth]{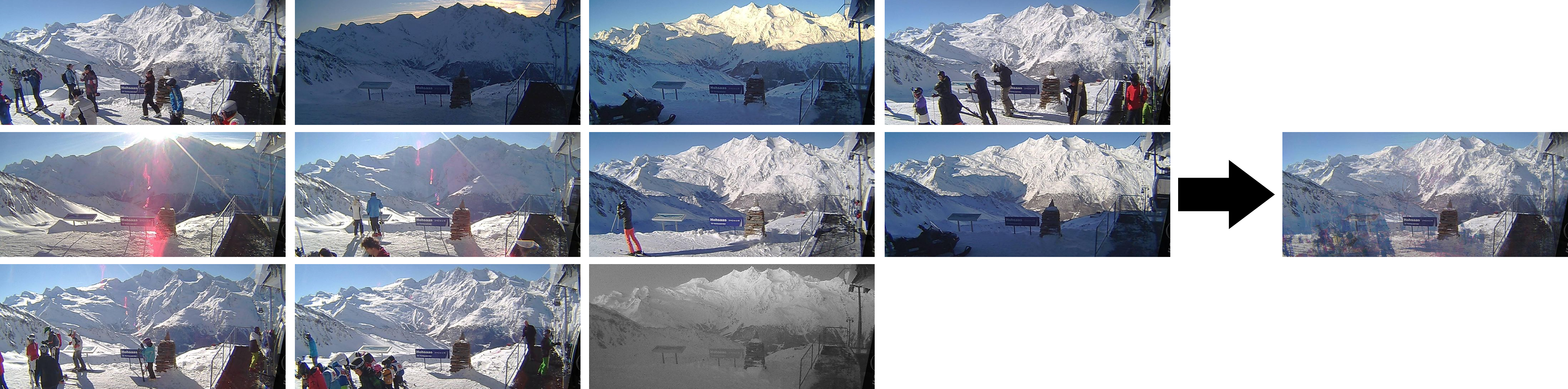}}
\caption{An example of a Daily Median Image (right) performed on 11 daily images (left).}
\label{fig:DMIExample}
\end{figure*}

Good weather images might suffer from challenging illumination conditions (such as solar glares and shadows) and moving obstacles (such as clouds and persons in front of the webcam). At the same time, snow cover changes slowly over time, so that one measurement per day is sufficient. Therefore, we aggregated the images collected by a webcam in a day, to obtain a single representative image to be used for further analysis. We adopted a simple median aggregation algorithm, which can deal with images taken in different conditions, removing transient occlusions and glares. Given $N$ good weather daily images $I_1,\dots,I_N$, we define the Daily Median Image ($DMI$) as
$$DMI(x,y) = med\{I_1(x,y),I_2(x,y),\dots,I_N(x,y)\},$$
where $med\{\cdot\}$ denotes the median operator, which is applied along the temporal dimension. Figure~\ref{fig:DMIExample} shows an example of a DMI generated by aggregating $11$ images. The aggregation attenuates the different illumination conditions and removes the persons standing in front of the webcam partially covering the mountain.

\begin{figure}
\centerline{\includegraphics[width=0.5\textwidth]{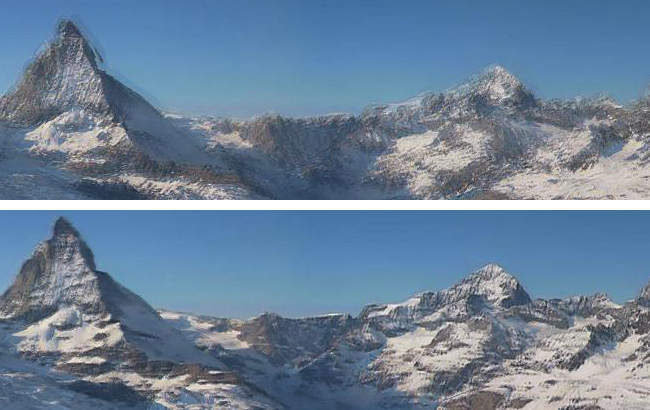}}
\caption{An example of a DMI performed without (top) and  with image registration (bottom).}
\label{fig:DMIPosAdjust}
\end{figure}

A challenging factor in the aggregation of the daily images lies in the fact that it is common for the webcam orientation to slightly vary during the day. This phenomenon might occur due to strong winds. The DMI of a webcam suffering from temporal jittering results in a blurry image, unsuitable for further analysis. To handle this issue, we performed image registration with respect to the reference frame of the first image. A global offset is computed by means of the cross-correlation between the two skyline edge maps. Each image is compensated by this offset before computing the DMI.
Figure~\ref{fig:DMIPosAdjust} shows an example DMI obtained without (top) and with image registration (bottom). The benefit obtained by using DMI in estimating snow cover is quantitatively analyzed in Section~\ref{sec:snow_cover}.

\section{Estimating Snow Cover}\label{sec:snow_cover}
\begin{figure}
\centerline{\includegraphics[width=0.5\textwidth]{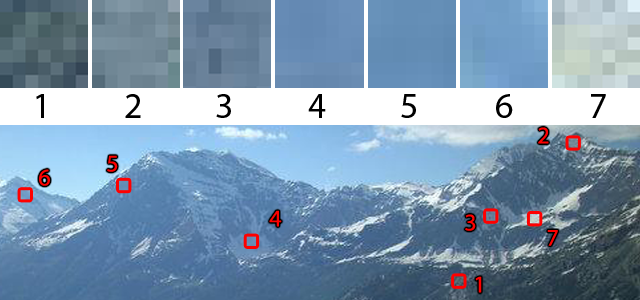}}
\caption{Several patterns of a webcam image. Patterns 1,5,6 represents terrain/vegetation area, while patterns 2,3,4,7 belong to the snow covered area.}
\label{fig:snowChallengingColor}
\end{figure}
Once a valid image that contains a mountain slope is retrieved, the area corresponding to the mountain surface must be analyzed and divided into snow and non snow areas. Although the segmentation of a mountain picture into snow and non snow areas is often a simple task for a human eye, it represents a challenging computer vision problem.
As an example, Figure~\ref{fig:snowChallengingColor} shows several $7\times7$ pixel patches extracted from a webcam image. If someone was asked to classify these patches as snow or terrain areas without looking at the image in the lower part of the figure, one would probably state that the first 3 patterns represent terrain, and the last 4 correspond to snow areas. Looking at the whole image, though, it would be possible to notice that, counter-intuitively, the patches n.2 and n.3 correspond to snow covered areas, whereas patches n.5 and n.6 are extracted from terrain areas. This example shows how the pixel-level snow classification heavily depends on the context of the image, and not only on single pixel intensities.

\subsection{Identifying snow covered area}
In this section we illustrate and evaluate approaches for pixel-level snow classification proposed in the literature. All listed methods adapt in an implicit or explicit way to different illumination conditions: threshold derived from statistical analysis of the pixel intensities~\cite{salvatori2011snow}, empirically defined color bands~\cite{hinkler2002automatic}, and probabilistic model fitting~\cite{rufenacht2014temporally}. Even so, all of them classify pixels as ``snow'' or ``non snow'' considering exclusively their intensity values. Conversely, given the challenging nature of the task, we study the benefits obtained by a supervised learning algorithm that considers also the context of each pixel.

Let $I$ denote the input image and $M$ the binary mask having the same size as $I$, where $M(x,y) = 1$ indicates that the corresponding pixel of the image belongs to the mountain area, or $M(x,y) = 0$ otherwise. The snow cover estimation is performed by a pixel-level binary classifier that, given $I$ and $M$ as input, produces a snow mask $S$ that assigns to each pixel a binary label denoting the presence of snow.

As a baseline, we consider a na\"{\i}ve method, henceforth called \textbf{Fixed Threshold}, which applies a simple threshold to a grayscale image, assuming that snow pixels are brighter.  Given an input grayscale image $G$ and a threshold value $\bar{t}$, the resulting snow mask is defined as:
$$
S(x,y) = \left\{\begin{matrix}
1 & \mbox{ }if\mbox{ }G(x,y) \geq \bar{t}\\ 
0 & otherwise
\end{matrix}\right..
$$

The methods for pixel-grained snow classification evaluated in this work include:

\textbf{Snow-noSnow}: Salvatori et al.~\cite{salvatori2011snow} propose a pixel level snow classifier called \emph{Snow-noSnow}. It is based on the analysis of the blue component of an RGB image, because the snow surface presents higher reflectance values in the blue wavelength range. The authors claim that in $90\%$ of the cases the histogram of any RGB component of a mountain image is shaped as a bimodal distribution. Let $B$ denote the blue component of the image normalized in the range $[0,255]$ and $BH$ the histogram of the intensity values of $\{B(x,y) | M(x,y) = 1\}$ where $M$ denotes the mountain area mask. The classifier of~\cite{salvatori2011snow} applies a threshold to each pixel of the blue component:
$$
S(x,y) = \left\{\begin{matrix}
1 & \mbox{ }if\mbox{ }B(x,y) \geq t\\ 
0 & otherwise
\end{matrix}\right.,
$$
where $t$ is equal to the first local minimum of $BH$ greater then $\bar{t}$, or $t = \bar{t}$ if such local minimum does not exist. The parameter $\bar{t}$ represents the lowest empiric intensity value of a snow pixel.

\textbf{RGB Normalized Difference Snow Index (RGBNDSI)}: Hinkler et al.~\cite{hinkler2002automatic} describe a classifier that applies a threshold not on a single color band, but on an empirically derived band, called RGBNDSI. The idea is to find a fictitious band, which is related to the Mid-Infrared (MIR) band used for Normalized Difference Snow Index calculation. Such index is used for the snow cover analysis of satellite imagery~\cite{salomonson2004estimating}. Let $R$, $G$ and $B$ denote, respectively, the three components of a true color image and let:
$$RGB = \frac{(R+G+B)}{3},$$
$$RGB_{high} = \frac{B^3}{R^3}G^3,$$
$$\tau = 200(a(avg(RGB_{high}))+b),$$
$$MIR_{replacement} = \frac{\tau^4max(RGB(x,y))}{RGB^4},$$
where $MIR_{replacement}$ is an empirical approximation of the MIR band, $\tau$ is an index of the brightness of the image and $RGB_{high}$ is an empirically derived matrix. The authors state that $\tau$ can be expressed as the mean of $RGB_{high}$, but a further linear transformation is applied to improve the performance in case of a large fraction of dark pixels. The values of $a$ and $b$ are derived by the authors for the specific camera used in the experiments, thus can not be applied to our context. For this reason, as suggested in~\cite{hinkler2002automatic}, we set $\tau = avg(RGB_{high})$. Finally, the derived color band to be thresholded is defined as
$$RGBNDSI = \frac{RGB-MIR_{replacement}}{RGB+MIR_{replacement}},$$
and the estimated snow mask is:
$$
S(x,y) = \left\{\begin{matrix}
1 & \mbox{ }if\mbox{ }RGBNDSI(x,y) \geq \bar{t}\\ 
0 & otherwise
\end{matrix}\right..
$$
Once again, the threshold value $\bar{t}$ is derived empirically. To this end, a statistical threshold selection method proposed by Salvatori et al.~\cite{salvatori2011snow} can be applied. The RGBNDSI method is an extension of the Snow-noSnow method, which replaces the blue band with an empirically derived one.

\textbf{Gaussian Mixture Model (GMM)}: R{\"u}fenacht et al.~\cite{rufenacht2014temporally} propose a snow classifier based on a GMM, where all the pixels to be classified are considered points in a 3 dimensional color space. A Gaussian mixture distribution with $k \geq 2$ components is fitted to the set of points $\{I(i,j) | M(x,y) = 1\}$. The Gaussian component with the highest mean intensity value is considered as the snow component, whereas all the others are deemed non-snow components. Each pixel is then labeled as snow, if its probability to belong to the snow component $p(x,y)$ is higher than a given threshold $\bar{t}$:
$$
S(x,y) = \left\{\begin{matrix}
1 & \mbox{ }if\mbox{ }p(x,y) \geq \bar{t}\\ 
0 & otherwise
\end{matrix}\right..
$$

\textbf{Supervised Learning Snow Classifiers}: in addition to the methods previously proposed in the literature, we considered supervised learning methods that, differently from the traditional approaches, consider also the context of every pixel.
For each pixel, a feature vector of $33$ elements is built and fed as input to a binary classifier. Given an image $I$, represented with a $3$ dimensional color space, the feature vector of each pixel $\{(x,y) | M(x,y) = 1\}$ is obtained as the concatenation of $3$ feature vectors, one for each color band $I^k$, $k = 1,2,3$. The $11$ elements feature vector of each color band includes: $9$ values for the pixel intensities contained in the $3\times3$ neighborhood of the analyzed pixel, 1 value representing the global intensity, and 1 for the local intensity. The global intensity is defined as the average intensity of all the pixels representing the mountain area, i.e., $avg(\{I^k(x_i,y_i) | M(x_i,y_i) = 1\} )$. The local intensity is the average intensity of the pixels within the mountain area, defined as $avg(\{I^k(x_i,y_i) | M(x_i,y_i) = 1 \land \| (x,y)-(x_i,y_i) \| \leq \bar{d} \} )$. The extent of the neighborhood is conveyed by the radius $\bar{d}$, which was set to $15$ pixels.
We evaluated this approach feeding the feature vectors to three supervised learning classifiers: \textbf{Support Vector Machine~(SVM)}, \textbf{Random Forest (RF)} and \textbf{Logistic Regression (LR)}.

\subsection{Snow mask post-processing}\label{sec:snow_postprocessing}
As mentioned before, it is common for a webcam to face bad weather conditions. If all the daily images are affected by low visibility it is not possible to produce the Daily Median Image (DMI) and to estimate the snow cover. Also, if the DMI is generated with few images, it can still suffer from solar glares and occlusions. In order to robustly estimate snow cover, it is possible to exploit the fact that such phenomenon varies slowly in time and that the neighborhood pixels are likely to belong to the same class (``snow'' or ``non-snow''). To this end, a post-processing method is proposed, which allows us to estimate the snow cover also for the days where no input data is available (due to missing data from the webcam or when all images are taken in bad weather conditions). Let $S_i, i=1,\dots,D,$ denote the snow mask of the $i$-th day, given a number $N$ of estimated daily snow masks observed in an interval of $D \geq N$ days. We obtain all missing snow masks by linear interpolation. For each day $i$, such that $S_i$ is missing, we consider the closest available masks, i.e. $S_{i-k_1}$ and $S_{i+k_2}$:
$$S_i(x,y) = \frac{k_1}{k_1+k_2}S_{i-k_1}(x,y) + \frac{k_2}{k_1+k_2}S_{i+k_2}(x,y).$$
Once snow masks are computed for each day, we apply a \textbf{median filter in the spatio-temporal domain} to each pixel of each mask, defined as
$$S_{i}^{s,t}(x,y) = med\{S_{i-t}(x-s,y-s), \dots , S_{i+t}(x+s,y+s)\},$$
where $med\{\cdot\}$ denotes the median operator, $s$ and $t$ are respectively the extent of the spatial and temporal window.

\subsection{Performance Evaluation}\label{sec:snow_performance}

\begin{figure}
\centerline{\includegraphics[width=0.5\textwidth]{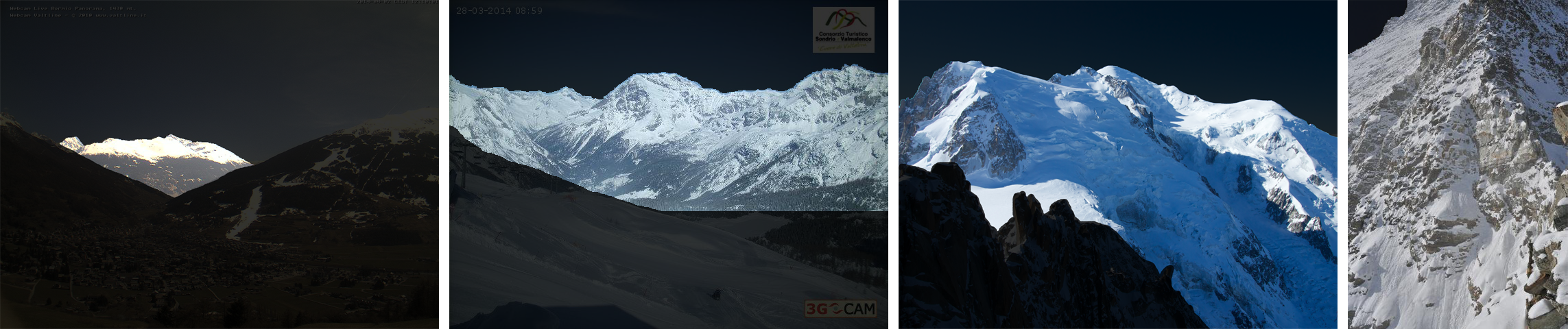}}
\caption{Sample images of the four different datasets (from left to right: \emph{Webcam 1}, \emph{Webcam 2}, \emph{UGC Photos}, \emph{PermaSense}).}
\label{fig:datasetExamples}
\end{figure}

\begin{table*}[t]
\begin{center}
\caption{Description of the datasets used for snow cover estimation experimental study.}
\label{tab:datasetDescription}
\begin{tabular}{|l|l|l|l|l|}
\hline
	Dataset
&	Description
&	Location
&	\# image
&	\# labeled images
\\ \hline
	\multicolumn{1}{|l|}{\multirow{2}{*}{Webcams}}
&	Webcam \#1: Single mountain, well defined snow line
&	Bormio, Italy
&	343
&	10
\\ 
	\cline{2-5}
	\multicolumn{1}{|l|}{}
&	Webcam \#2: Plural mountain peaks, snow at different altitudes
&	Valmalenco, Italy
&	338
&	10
\\ \hline
	PermaSense
&	Webcam framing a small portion of Matterhorn mountain & Switzerland
&	2491
&	19
\\ \hline
	UGC Photos
&	Random sample of crawled mountain photographs
&	Italy-Switzerland border
&	20
&	20
\\ \hline
\end{tabular}
\end{center}
\end{table*}

\textbf{Datasets:} to evaluate the performance of the snow cover estimation methods we considered $3$ different datasets. The \emph{Webcams} dataset comprises the images collected from two publicly available webcams placed in proximity of ground meteorological stations. This allowed us to have a reliable source of data for the study of the consistency of the snow estimations with respect to other measurements, such as air temperature. The \emph{PermaSense} dataset was collected by the PermaSense project\footnote{http://www.permasense.ch} at the Matterhorn field site and used in~\cite{rufenacht2014temporally}. The \emph{UGC Photos} dataset is a subset of randomly extracted mountain photographs crawled from Flickr, as described in Section~\ref{sec:photos_acquisition}. Figure~\ref{fig:datasetExamples} shows a sample image from each dataset, highlighting with the opacity the region of interest (i.e. the binary mask $M$), while Table~\ref{tab:datasetDescription} reports the detailed information about the datasets. For each dataset, a subset of the images uniformly distributed over time was selected, and for each image the groundtruth snow mask was created by manually tagging all the mountain area pixels as ``snow'' or ``non-snow''. Each image of the \emph{UGC Photos} dataset was included in the labeled image set. A total of $7M$ pixels contained in 59 images were manually labeled. Each dataset has its own specific characteristics and is studied separately.

In order to normalize the testing conditions, all the input images were downsampled so that at least one of the dimensions reached a fixed maximum value ($\bar{w}$ and $\bar{h}$ respectively). A scale factor $k = min(1, max(\frac{\bar{w}}{w},\frac{\bar{h}}{h}))$ was applied to each image, where $w$ and $h$ are respectively the width and the height of the image. In our implementation we set $\bar{w} = 640$ and $\bar{h} = 480$.

For each dataset, we defined $P_i$ as the collection of all the pixels that were assigned a label snow/non-snow. $P_i$ was split in a subset of $80k$ samples forming the test set, and the remaining samples were assigned to the training set.
In order to evaluate the ability of the supervised learning classifiers to adapt to different imaging conditions, they were trained using the data equally distributed from all the datasets.

\textbf{Results:}

\begin{figure*}
\centering
\begin{subfigure}{0.32\textwidth}
	\centerline{\includegraphics[width=1.0\textwidth]{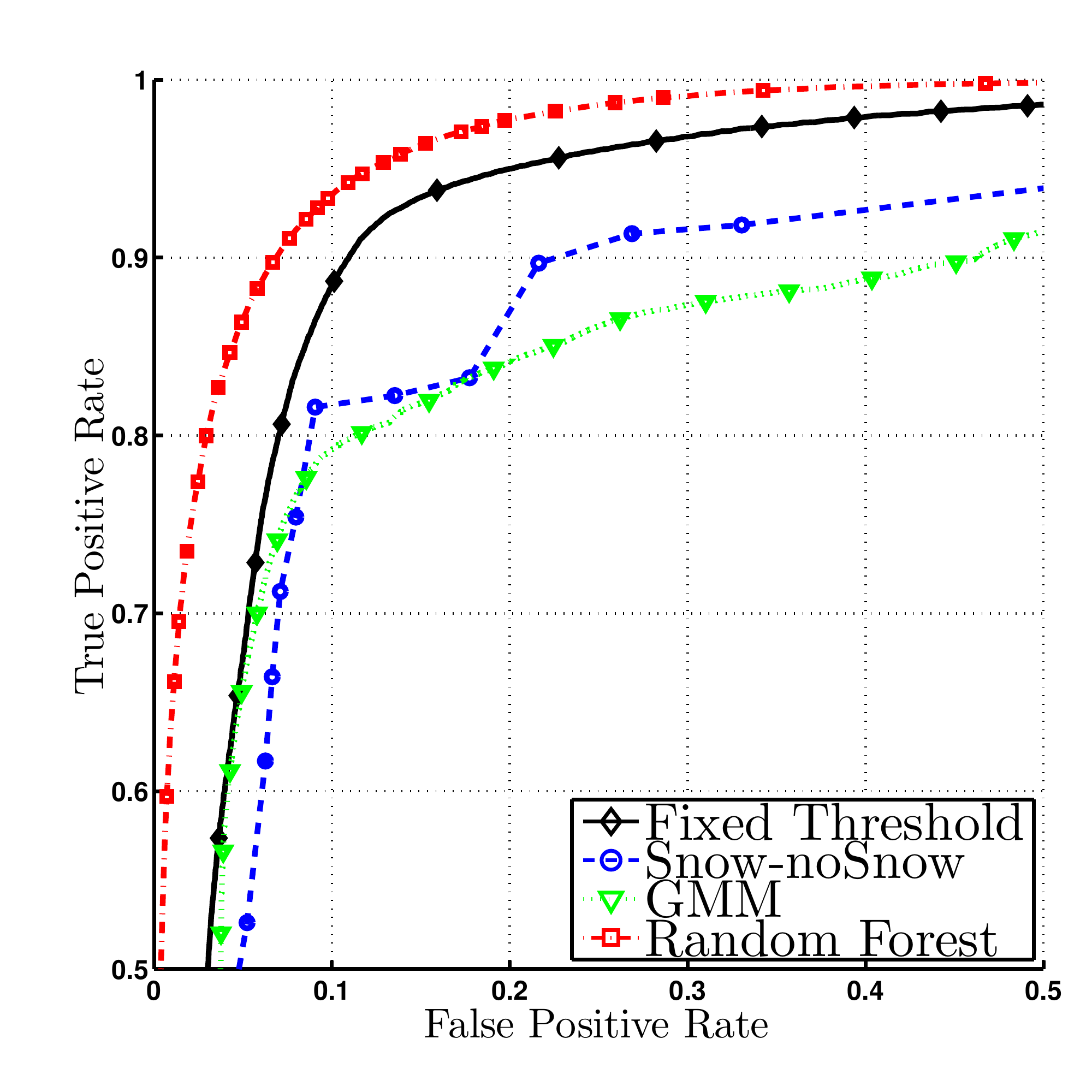}}
	\caption{\emph{Webcams} dataset}
	\label{fig:rocWebcams}
\end{subfigure}           
\begin{subfigure}{0.32\textwidth}
	\centerline{\includegraphics[width=1.0\textwidth]{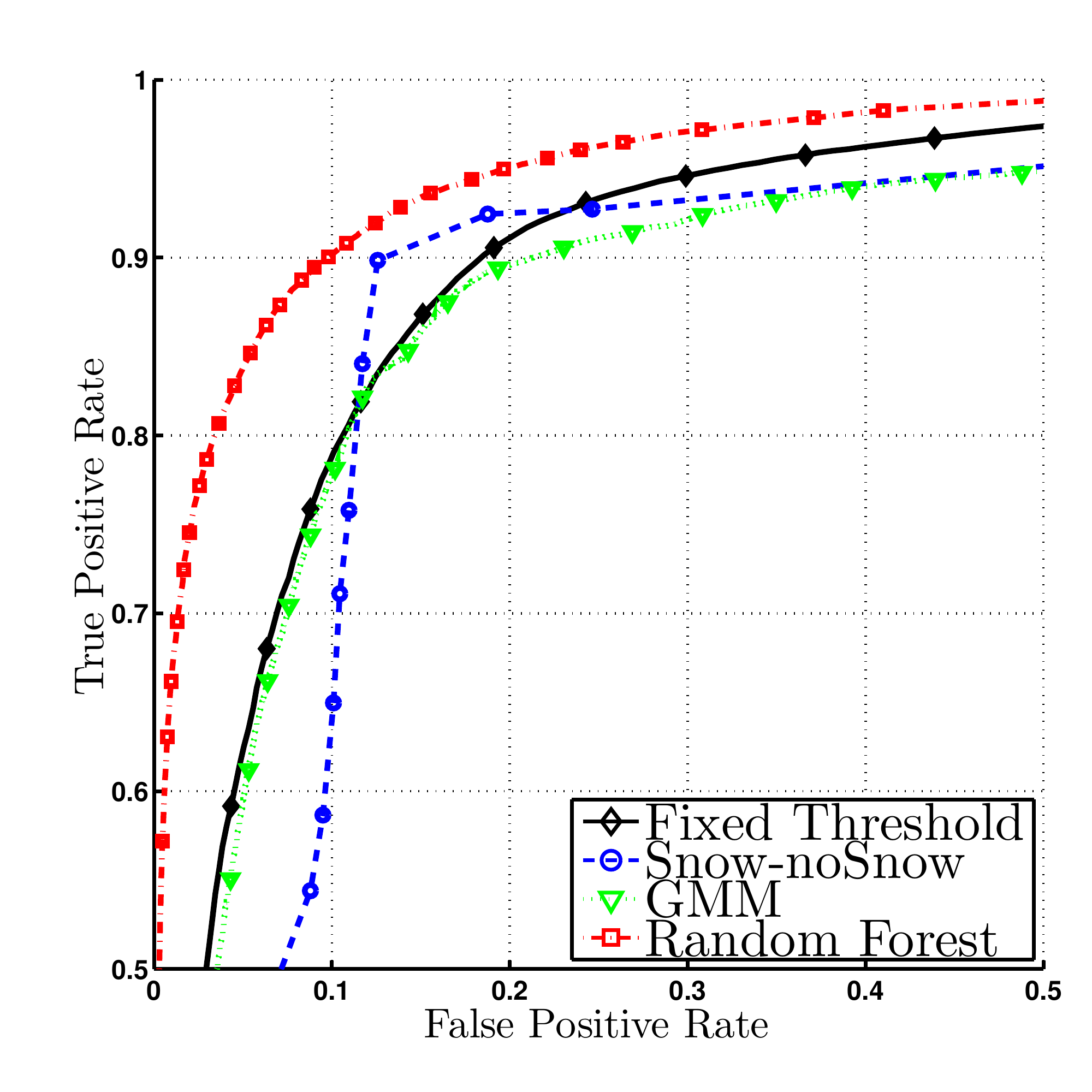}}
	\caption{\emph{PermaSense} dataset}                 
	\label{fig:rocPersen}
\end{subfigure}       
\begin{subfigure}{0.32\textwidth}
	\centerline{\includegraphics[width=1.0\textwidth]{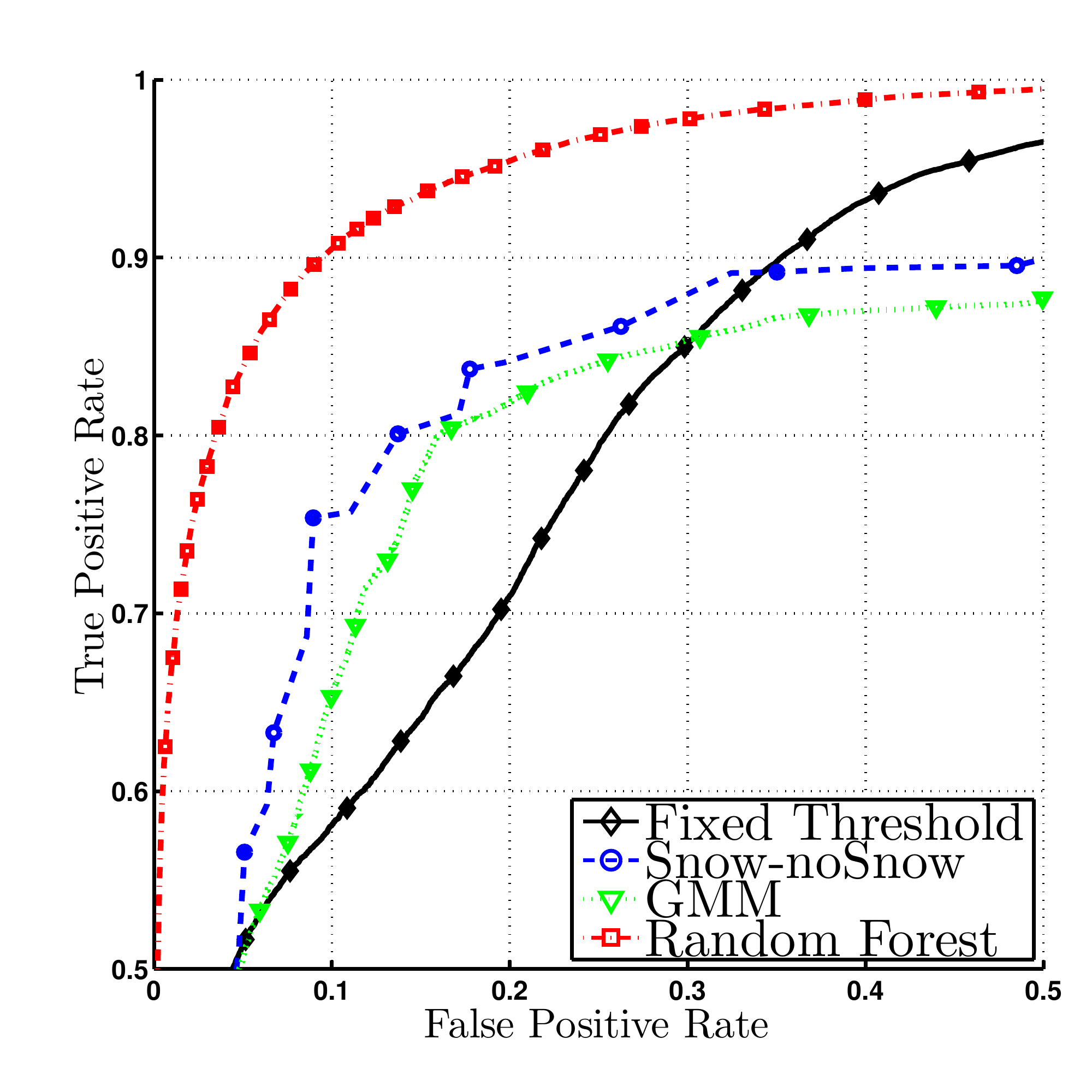}}
	\caption{\emph{UGC Photos} dataset}
	\label{fig:rocPhotos}
\end{subfigure} 
\caption{ROC curves obtained in the analyzed datasets by the different snow classifiers.}
\label{fig:rocs}
\end{figure*}

Figure~\ref{fig:rocWebcams} shows the ROC curve obtained on the \emph{Webcams} dataset. We report results for the \emph{Fixed Threshold} baseline, the \emph{Snow-noSnow} classifier, and the \emph{GMM} classifier with $3$ Gaussian components.
We also include the results obtained by the \emph{Random Forest (RF)} classifier - as the best performing supervised learning method - which was trained on equally balanced data from all datasets.
The number of trees of the RF is $50$, all the variables are selected for each decision split.
Dealing with images taken from the same webcam, the \emph{Fixed Threshold} method adapts efficiently to the common illumination factors and performs better than the other non-supervised methods (including \emph{RGBNDSI}, not shown to avoid cluttering the figure).
The \emph{RF} method dominates the others, showing the improvement obtained by exploiting the pixel context.

Table~\ref{tab:TPR10Webcams} shows the True Positive Rate (TPR) obtained by all classifiers when keeping the False Positive Rate (FPR) fixed at $0.1$. The GMM method was evaluated with $2$ and $3$ Gaussian components.
The first column contains the TPR obtained without using Daily Median Images (DMIs), as described in Section~\ref{sec:webcams_processing}, and without the spatio-temporal median filter.
In this case, the image with the highest highest skyline visibility score (described in the section~\ref{sec:webcams_acquisition}) is used as the representative of each day.
The second column shows the TPR with the DMI.
The third column specifies the TPR obtained with both the spatio-temporal median filter and the DMI.
The use of the DMI improves the performance of all methods, while the spatio-temporal median filtering has a positive impact only on those methods obtaining low TPR (\emph{GMM} methods) and a negative impact on more accurate methods (\emph{RF}, \emph{SVM}, \emph{LR}, \emph{RGBNDSI}, \emph{Snow-noSnow}, \emph{Fixed Threshold}), due to over-smoothing. Namely $S^{s=1,t=1}_i$ was computed, but the same trend has been observed for different values of $s$ and $t$.


Figure~\ref{fig:rocPersen} shows the ROC curves obtained within the \emph{PermaSense} dataset. Since the images in the \emph{PermaSense} dataset represent a fragment of the \emph{Matterhorn}, it was not possible to apply the skyline visibility score described in the Section~\ref{sec:webcams_acquisition}. Hence, we used the statistical methods for the bad weather image filtering based on color analysis, proposed by the authors of the \emph{GMM} method~\cite{rufenacht2014temporally}.
Table~\ref{tab:TPR10Persen} shows the TPR obtained for the \emph{PermaSense} dataset keeping the FPR fixed at $0.1$. All the classifiers, with the exception of the \emph{RF}, \emph{SVM}, \emph{LR} benefit from the spatio-temporal median filtering. Analogously to the \emph{Webcams} dataset, all the methods benefit from the use of Daily Median Images.  

\begin{table}[h]
    \centering    
    \begin{minipage}{1.0\linewidth}
	  \caption{TPR obtained in the \emph{Webcams} dataset by the different snow classifiers.}
	  \label{tab:TPR10Webcams}
      \centering
\begin{tabular}{l|l|l|l|}
\cline{2-4}
	& \begin{tabular}[x]{@{}c@{}}without Median\\without DMI\end{tabular}
	& \begin{tabular}[x]{@{}c@{}}without Median\\with DMI\end{tabular}
	& \begin{tabular}[x]{@{}c@{}}with Median\\with DMI\end{tabular}
\\ \hline
\multicolumn{1}{|l|}{Random Forest}            & 79.7   & \textbf{93.5} (+13.8)          & \textbf{91.6} (+11.9)
\\ \hline
\multicolumn{1}{|l|}{SVM}             & \textbf{81.8}   & 92.6 (+10.8)          & \textbf{91.6} (+9.8)
\\ \hline
\multicolumn{1}{|l|}{Linear Regression} & 78.3 & 89.7 (+11.4)      & 88.8 (+10.5)
\\ \hline
\multicolumn{1}{|l|}{GMM3}            & 73.7   & 79.2 (+5.5)      & 82.7 (+9.0) 
\\ \hline
\multicolumn{1}{|l|}{GMM2}         	  & 79.9   & 80.2 (+0.3)      & 83.3 (+3.4) 
\\ \hline
\multicolumn{1}{|l|}{RGBNDSI}         & 72.9   & 87.1 (+14.2)     & 87.0 (+14.1)
\\ \hline
\multicolumn{1}{|l|}{Snow-noSnow}     & 68.6   & 81.7 (+13.1)     & 80.5 (+11.9)
\\ \hline
\multicolumn{1}{|l|}{Fixed Threshold} & 76.0   & 88.4 (+12.4)     & 87.6 (+11.6)
\\ \hline
\end{tabular}
    \end{minipage}%

\vspace{\baselineskip}
    
    \begin{minipage}{1.0\linewidth}
      \centering
\caption{TPR obtained in the \emph{PermaSense} dataset by the different snow classifiers.}
\label{tab:TPR10Persen}
\begin{tabular}{l|l|l|l|}
\cline{2-4}
	& \begin{tabular}[x]{@{}c@{}}without Median\\without DMI\end{tabular}
	& \begin{tabular}[x]{@{}c@{}}without Median\\with DMI\end{tabular}
	& \begin{tabular}[x]{@{}c@{}}with Median\\with DMI\end{tabular}
\\ \hline
\multicolumn{1}{|l|}{Random Forest}             & 83.8 & \textbf{90.2} (+6.4)          & \textbf{89.2} (+5.4)	
\\ \hline
\multicolumn{1}{|l|}{SVM}             & 84.7   & 89.2 (+4.5)      & 88.7 (+4.0) 
\\ \hline
\multicolumn{1}{|l|}{Linear Regression} & \textbf{87.0} & 87.9 (+0.9)      & 87.2 (+0.2) 
\\ \hline
\multicolumn{1}{|l|}{GMM3}            & 70.6   & 77.8 (+7.2)      & 83.2 (+12.6)  
\\ \hline
\multicolumn{1}{|l|}{GMM2}            & 60.2   & 78.6 (+18.4)     & 84.7 (+24.5)  
\\ \hline
\multicolumn{1}{|l|}{RGBNDSI}         & 73.6   & 78.3 (+4.7)      & 84.7 (+11.1) 
\\ \hline
\multicolumn{1}{|l|}{Snow-noSnow}     & 59.7   & 64.1 (+4.4)      & 87.1 (+27.4) 
\\ \hline
\multicolumn{1}{|l|}{Fixed Threshold} & 66.5   & 78.8 (+12.3)     & 82.3 (+15.8)
\\ \hline
\end{tabular}
    \end{minipage} 
\end{table}
         
Figure~\ref{fig:rocPhotos} depicts the ROC curves obtained on the \emph{UGC Photos} dataset. The major difference with respect to the other datasets is the fact that the \emph{Fixed Threshold} method is dominated by all the others. These results are as expected, as the photographs were taken in different locations, with different cameras and different conditions. The \emph{Fixed Threshold} method is not able to find a threshold value that is suitable for all images, while the other methods are more capable of adapting to varying conditions. In this case, we do not report the results as in Table~\ref{tab:TPR10Webcams} because median filtering and DMI can not be applied to this dataset that contains spatially and temporally independent images.

\subsection{Computing the Snow Cover Index}\label{sec:snow_cover_index}
The pipeline described in this work produces a pixel-wise snow cover estimation from photographs and webcam images, along with all the metadata necessary for feeding environmental models, such as GPS position, camera orientation, and mountain peak alignment. Although environmental modeling is outside the scope of this work, we performed a case study to show the consistency of the snow cover estimations with other environmental variables. We monitored the snow melting process during a two month period for a mountain captured by a webcam in the Italian Alps. In particular, we studied the elevation of the snow line, i.e. the minimum elevation at which snow is present.

\begin{figure*}
	\centerline{\includegraphics[width=\textwidth]{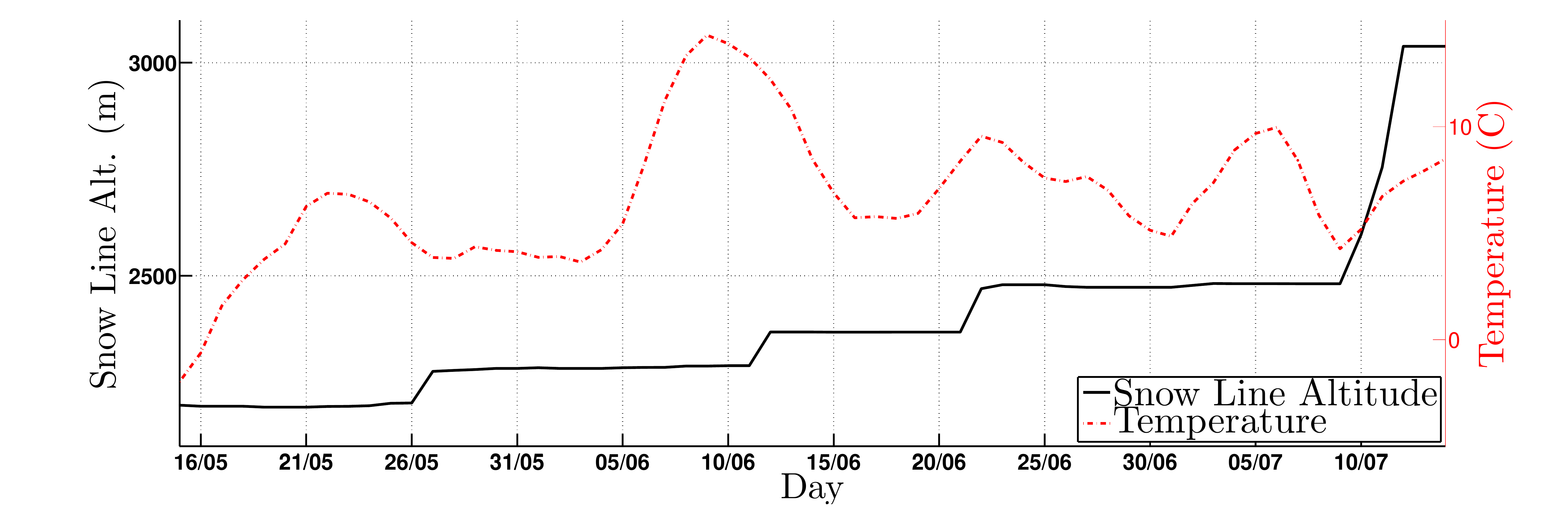}}
	\caption{The snow level altitude and the temperature trends during the observation period.}                                
	\label{fig:bormioSnowLevel}
\end{figure*} 

Given a snow mask $S$, the area of the image covered by the mountain was split into $N$ horizontal bands. Because the image was aligned with the rendered terrain view, we were able to estimate the altitude of each image pixel. Let $A_i$ denote the altitude of the lowest pixel of the $i$-th horizontal band (where $A_1$ and $A_N$ corresponds respectively to the lowest and highest altitude bands). We define Snow Vector Index (SVI) a vector of $N$ elements, where $SVI_i$ is a number in the range $[0,1]$ that defines the fraction of the $i$-th horizontal band containing snow pixels. In other words, the SVI can be seen as the snow cover percentage at different altitude levels. Given an SVI, we estimate the snow line altitude $L$ as $L = A_k + SVI_k(A_{k+1}-A_k)$, for a value of $k$ such that $SVI_{k-1} < \bar{s}$ and $SVI_{k},\dots,SVI_{N} \geq \bar{s}$ where $\bar{s}$ is the threshold that defines the maximum negligible snow cover percentage. 

We studied the snow line altitude dynamics for one of the webcams used in the \emph{Webcams} dataset (see Figure~\ref{fig:datasetExamples}~left). We acquired 40k images during a two month period going from May 15th to July 14th. For 49 days (out of the 61 days of the monitored period) at least one good weather image was retrieved and the corresponding DMI was generated. Then, a snow mask was extracted for each DMI and the missing day snow masks estimated by the interpolation described in Section~\ref{sec:snow_postprocessing}. Finally, the snow line altitude was estimated for each observed day.
Figure~\ref{fig:bormioSnowLevel} shows the trend of the snow level altitude (smoothed by a median filter with the window size equal to $4$ days), along with the air temperature registered by a nearby ground station. It can be observed that the snow melting process was characterized by four occurrences when the snow level altitude increased abruptly. This behavior is correlated with the four temperature peaks observed by the meteorological station.
This example shows a possible application of the snow cover estimation based on public visual content, and confirms the consistency of the proposed methods and metrics.

\section{Conclusions and Future Work}\label{sec:conclusions}
In this paper we addressed the challenge of estimating snow cover from publicly available images. We introduced methods for the automatic acquisition of relevant mountain photographs and webcam images. Then, we discussed techniques for the estimation of photograph orientation and for the mountain peak identification. Furthermore, we compared different methods for the estimation of snow cover from a mountain image. Finally, we reported a case study, in which we monitored the elevation of the snow line of a mountain to prove the consistency of the snow estimations from public contents.

The future work will investigate the applicability of the proposed techniques to several environment-related problems. A notable example is the calculation of Snow Water Equivalent (SWE): in snow covered basins, the accurate estimate of the SWE, i.e., the amount of water contained in the snow pack, is key to improve the anticipation capability of decision making in operational flood control, water supply planning, and water resources management.
SWE is usually estimated by spatial interpolation of ground-sensed point measurements of snow depth and density conditioned on the snow-covered area retrieved by satellite images processing. Since the middle of the 1960s, a number of satellite-derived snow products have been available to complement low-density snow monitoring networks, especially at inaccessible mountainous or high latitude regions. Satellite products, however, suffer from some technical limitations that hinder their operational value in most alpine contexts: space-board passive microwave radiometers (e.g. AMSR-E) easily penetrate clouds and provide accurate estimation, but work on very coarse spatial resolution. Optical sensors (e.g. MODIS) generate high-resolution snow cover maps, yet cannot see the earth surface when clouds are present.

We argue that the reported results reveal the potential benefit of exploiting the user generated content as an additional source of information to be used in conjunction with ground and remote sensing, for producing high temporal (daily to weekly) and spatial resolution snow phenomena time series.


%




\ifCLASSOPTIONcaptionsoff
  \newpage
\fi



%
%
%
\bibliographystyle{IEEEtran}
\bibliography{IEEEabrv,snowwatch_ieeetom}

%

\begin{IEEEbiography}
[{\includegraphics[width=1in,height=1.25in,clip,keepaspectratio]{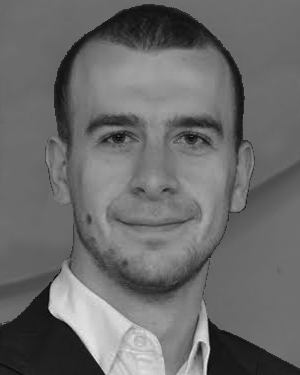}}]{Roman Fedorov}
received the M.Sc. (cum laude) degree in computer engineering from Politecnico di Milano, Milan, Italy, in 2013, where he is currently pursuing the Ph.D. degree in the Dipartimento di Elettronica e Informazione.

His research interests are in the areas of collective intelligence extraction from user-generated content and social data mining and analysis.	
\end{IEEEbiography}

\begin{IEEEbiography}
[{\includegraphics[width=1in,height=1.25in,clip,keepaspectratio]{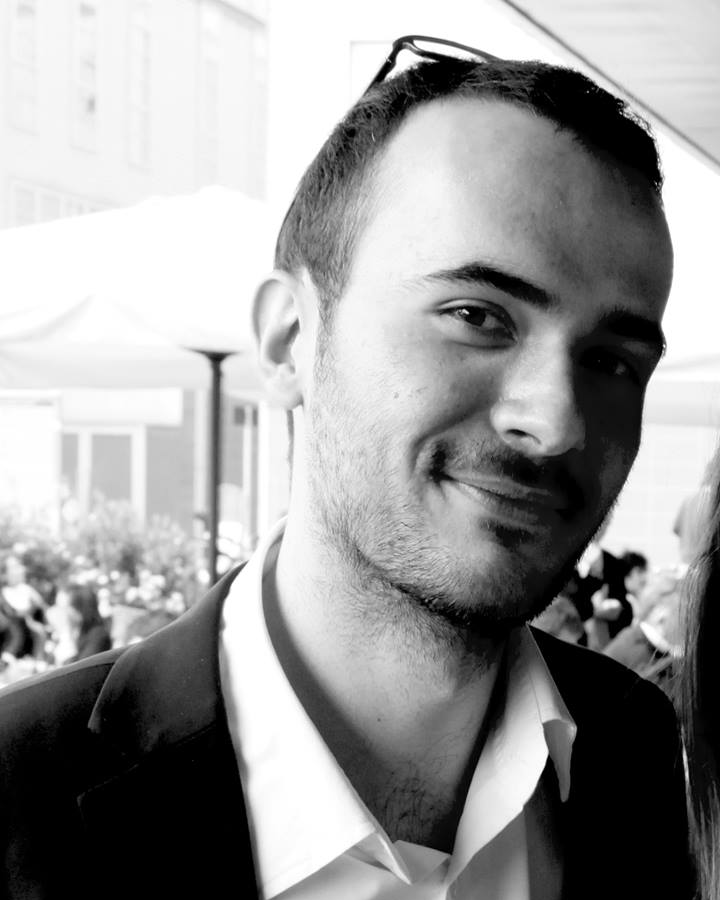}}]{Alessandro Camerada}
received the M.Sc. degree in computer engineering from Politecnico di Milano, Milan, Italy, in 2014.
\end{IEEEbiography}

\begin{IEEEbiography}
[{\includegraphics[width=1in,height=1.25in,clip,keepaspectratio]{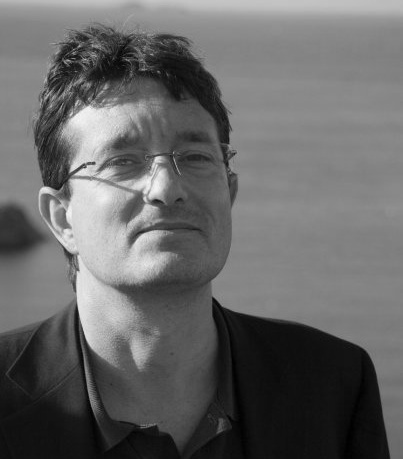}}]{Piero Fraternali}
is full professor 
of Web Technologies 
at DEIB, Politecnico di Milano, Italy. His main research interests concern software engineering, and methodologies, tools for Web application development, multimedia information retrieval and human computation. 
He is co-inventor of WebML, a model for the conceptual design of Web applications 
(US Patent 6,591,271, July 2003) 
and co-author of the Interaction Flow Modeling Language (IFML), the OMG standard for application interface modeling. 
He served as Program Chair of the International Conference on Web Engineering in 2004, Vice President of the Software Engineering Track of the WWW conference in 2005, General Chair of the International Conference on Web Engineering in 2007 and Area Chair of the WWW Conference in 2010.
\end{IEEEbiography}

\begin{IEEEbiography}
[{\includegraphics[width=1in,height=1.25in,clip,keepaspectratio]{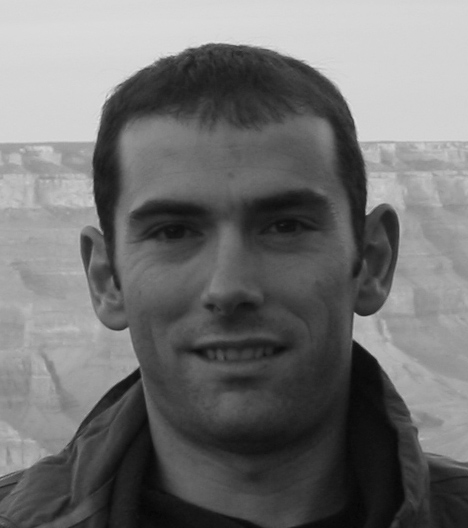}}]{Marco Tagliasacchi}
is currently Associate Professor at DEIB, Politecnico di Milano, Italy. He received the 
MS degree (2002) in Computer Engineering and the Ph.D. in Electrical Engineering and Computer Science (2006), both from Politecnico di Milano. 
His research interests include multimedia forensics, multimedia communications (visual sensor networks, coding, quality assessment) and information retrieval.
\end{IEEEbiography}








\end{document}